\documentclass[pra,aps,twocolumn,amsmath,amssymb,floatfix,reprint,footinbib,superscriptaddress]{revtex4-2}
\usepackage{graphics}      
\usepackage{graphicx}      
\usepackage{longtable}     
\usepackage{url}           
\usepackage{bm}    
\usepackage{braket}        
\usepackage{xcolor}
\usepackage{comment}
\usepackage{siunitx}
\usepackage[T1]{fontenc}
\usepackage[title]{appendix}
\usepackage{amsmath,amssymb,epsfig,color}
\usepackage{mathrsfs}
\usepackage{amsfonts}
\usepackage{dcolumn}
\usepackage{subfigure}
\usepackage{soul}
\usepackage{tabularx}
\usepackage[colorlinks,linkcolor=blue,citecolor=blue,hyperindex,bookmarks=false,pdfstartview=FitH]{hyperref}

\usepackage[ruled]{algorithm2e}  
\newcommand{\beq}{\begin{equation}}
\newcommand{\eeq}{\end{equation}}
\newcommand{\beqa}{\begin{eqnarray}}
\newcommand{\eeqa}{\end{eqnarray}}

\newcommand{\red}[1]{\textcolor{red}{#1}}

\newcommand{\nn}{\nonumber}
\newcommand{\be}{\begin{equation}}
\newcommand{\ee}{\end{equation}}
\newcommand{\bea}{\begin{eqnarray}}
\newcommand{\eea}{\end{eqnarray}}

\newcommand{\figpanel}[2]{\hyperref[#1]{\ref*{#1}(#2)}}
\newcommand{\figpanels}[3]{\hyperref[#1]{\ref*{#1}(#2)-(#3)}}
\newcommand{\figpanelNoPrefix}[2]{\hyperref[#1]{\ref*{#1}(#2)}}
\usepackage{mleftright} 

\begin{document}

\title {Giant atom with disorders: Effects from imperfect couplings}
\author{Muming Han}
\affiliation{Center for Joint Quantum Studies and Department of Physics, School of Science, Tianjin University, Tianjin 300072, China}
\author{Lingzhen Guo}
\email{lingzhen\_guo@tju.edu.cn}
\affiliation{Center for Joint Quantum Studies and Department of Physics, School of Science, Tianjin University, Tianjin 300072, China}

\begin{abstract}
The study of giant atoms goes beyond the local interaction paradigm in conventional quantum optics and predicts novel phenomena, such as oscillating bound states in the continuum (BICs) and decoherence-free interaction (DFI) that do not exist in small atoms, for some particular parameter settings of coupling positions and strengths. However, in the realistic experiments to implement giant-atom systems, there is always some level of disorder both in coupling positions and strengths.  In this work,  we investigate the effects of disorder on the phenomena related to giant atoms. We find that the giant-atom-related phenomena are robust to both disorders of coupling positions and strengths in the Markovian regime, but more sensitive to the disorder of coupling positions in the non-Markovian regime. Our work shows that, to observe the non-Markovian phenomenon such as (oscillating) BICs in giant-atom systems, more precision is needed to control the disorder of coupling positions than that of the coupling strengths in the experiments.
\end{abstract}
\date{\today}
\maketitle

\section{Introduction}\label{sec:level1}
In the standard framework of quantum optics, the local dipole-dipole interaction that treats the atom as a point-like model is a well-justified approximation as the size of atoms is usually much smaller than the wavelength of bosonic modes in the environment~\cite{Walls2008}. In recent years, the study of \textit{giant atoms}, that goes beyond the local dipole approximation representing a new paradigm and research frontier in quantum optics \cite{anton2021book,Gonzalez-Tudela2019,Calajo2019,Lorenzo2021sr} has drawn more and more attention. The artificial atoms made in the laboratory can be comparable to or much larger than the wavelength of bosonic modes in the environment, e.g., a superconducting transmon \cite{Koch2007pra} nonlocally coupled to surface acoustic waves (SAWs) \cite{Gustafsson2014science,Andersson2019np} or meandering microwave transmission line~\cite{Kannan2019,vadiraj2021pra} via multiple coupling points. The magnon-based hybrid system of a ferromagnetic spin ensemble interacting multiple times with the meandering waveguide can also be treated as a giant-atom system~\cite{Wang2022nc}.
Many novel phenomena absent in small-atom systems, e.g.,  frequency-dependent relaxation rates/Lamb shifts~\cite{Kockum2014pra}, \textit{decoherence-free interaction} (DFI) between giant atoms \cite{Kockum2018prl,carollo2020prr}, non-exponential spontaneous emission \cite{Guo2017,Andersson2019np} and persistently oscillating \textit{bound states in the continuum} (BICs)~\cite{guo2020prr,taylor2020pra}, arise for giant atoms due to the self-interference effects or non-Markovian time delays.

The study of giant atom(s) originally developed from the system of superconducting artificial atom(s) coupled to a linear waveguide~\cite{Kockum2014pra,Guo2017,sheremet2023rmp} has been  extended to the systems of giant atom(s) interacting with a structured environment, e.g., an array of coupled resonators/cavities \cite{zhao2020pra,lim2023pra,zhang2023pra,jia2023atomphoton,bag2023quantum} or linear waveguide with designed spatial coupling sequence \cite{wang2023realizing}. The giant atom(s) models were also proposed to be realized in a topological environment~\cite{cheng2022pra,vega2021pra,vega2023prr}, in a non-Hermitian environment~\cite{du2023giant}, in the two-dimensional environment with atomic matter wave~\cite{Gonzalez-Tudela2019} or in the optical regime with coupled Rydberg atoms \cite{chen2023giantatom}. The two-level giant atom has been extended to the three-level $\Lambda$-type or $\Delta$-type artificial atoms~\cite{zhu2022pra,Chen2022cp,gu2023correlated,gong2024arxiv}. The real-space environment that interacts with giant atoms has also been extended to the synthetic frequency dimension \cite{du2022prl}.
Giant atoms provide a new platform to study the chiral or nonreciprocal phenomena such as directed emission and chiral bound states  \cite{li2023tunable,wang2021prl} for chiral quantum optics \cite{wang2023realizing,Guimond2020,zhang2021prl,joshi2023prx,Kannan2023nat,Wang2022qst,du2022prl,vega2023prr}.
The entanglement dynamics in double-giant-atom waveguide-QED systems were studied~\cite{yin2023pra,yin2022pra}, and maximally entangled long-lived states of giant atoms can be generated by a resonant classical field  \cite{santos2023prl}.
The rich physics of giant atoms in the ultrastrong coupling regime~\cite{Ask2019,noachtar2022pra,zueco2022pra} or the multi-photon/phonon process~\cite{Guo2017,Guimond2017,gu2023correlated,cheng2023lp,arranz2021prr,pichler2016prl,wang2024arxiv} are of great interest in the recent studies.

\begin{figure}
\centering
\includegraphics[width=\linewidth]{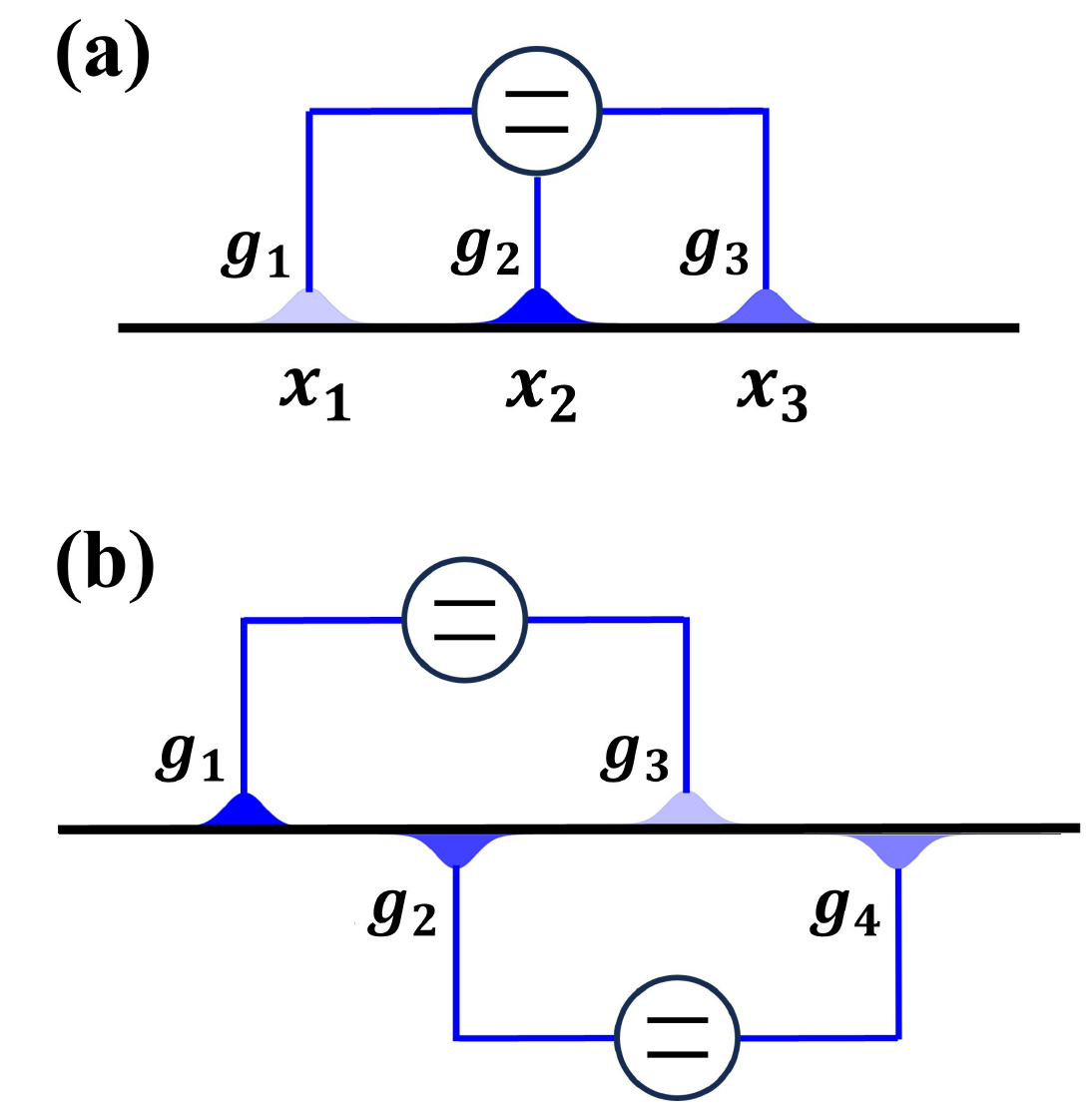}
\caption{Sketch of two-level giant atom(s) coupled to an open 1D waveguide at multiple $N$ coupling points $x_i$ with coupling strength $g_i$ : {\bf (a)} One giant atom with $N = 3$ coupling points; {\bf (b)} Two braided giant atoms with total $N = 4$ coupling points. In both figures, the disorders of coupling positions and coupling strengths are illustrated by the envelope and the transparency of the blue wave packets, respectively.}
\label{fig-model-2}
\end{figure}

Giant atoms also provide new possibilities to design functional quantum devices that are important for building scalable quantum networks, e.g., chiral single-photon emitter (absorber) \cite{Guimond2020,Kannan2023nat} or router \cite{Chen2022cp, Wang2021oe,wang2024arxiv}  based on the size-induced self-interference effects of giant atom \cite{Guo2017,zhu2022pra,Qiu2023,yin2022pra,du2023pra,lim2023pra,cheng2022pra,ask2022prl,arranz2021prr,pichler2016prl,yin2022pra}. The non-Markovianity due to the time delays among the legs of giant atoms can also be a valuable resource for quantum information processing \cite{hannes2017pans,Sletten2019} and has been proposed to tweeze (catch and release) the propagating field in the environment on demand~\cite{xu2024njp}.

However, in realistic experiments to implement giant atoms, there is always some level of disorder in both the coupling positions and the strengths. The effects of such disorders on giant atoms have not yet been explored. In this work, we investigate the effects of disorders on the basic giant-atom phenomena: DFI and BICs. As expected, both DFI and BICs are suppressed and deteriorated by the disorders of coupling positions and strengths. Interestingly, we find that both the DFI and BICs effects are suppressed by the two disorders basically in a similar manner in the Markovian regime, but are more sensitive to the position disorder in the non-Markovian regime.

Our paper is organized as follows. In Section~\ref{Model Hamitonian}, we introduce the model
Hamiltonian of giant atoms coupled to a waveguide via multiple coupling points with disordered coupling positions and strengths. 
In Section~\ref{sec-BIC}, we investigate the effects of disorders on the dark states and BICs for a non-Markovian giant atom based on the equation of motions (EOMs) for the giant atom.
In Section~\ref{sec-DFI}, we study the effects of disorders on the DFI of braided giant atoms based on the master equation from the SLH formalism~\cite{Kockum2018prl}.
In Section~\ref{sec-sumlook}, we summarize the results in the paper and outline the possible research directions in the future.

\begin{figure}[htp]
\centering
\includegraphics[width=\linewidth]{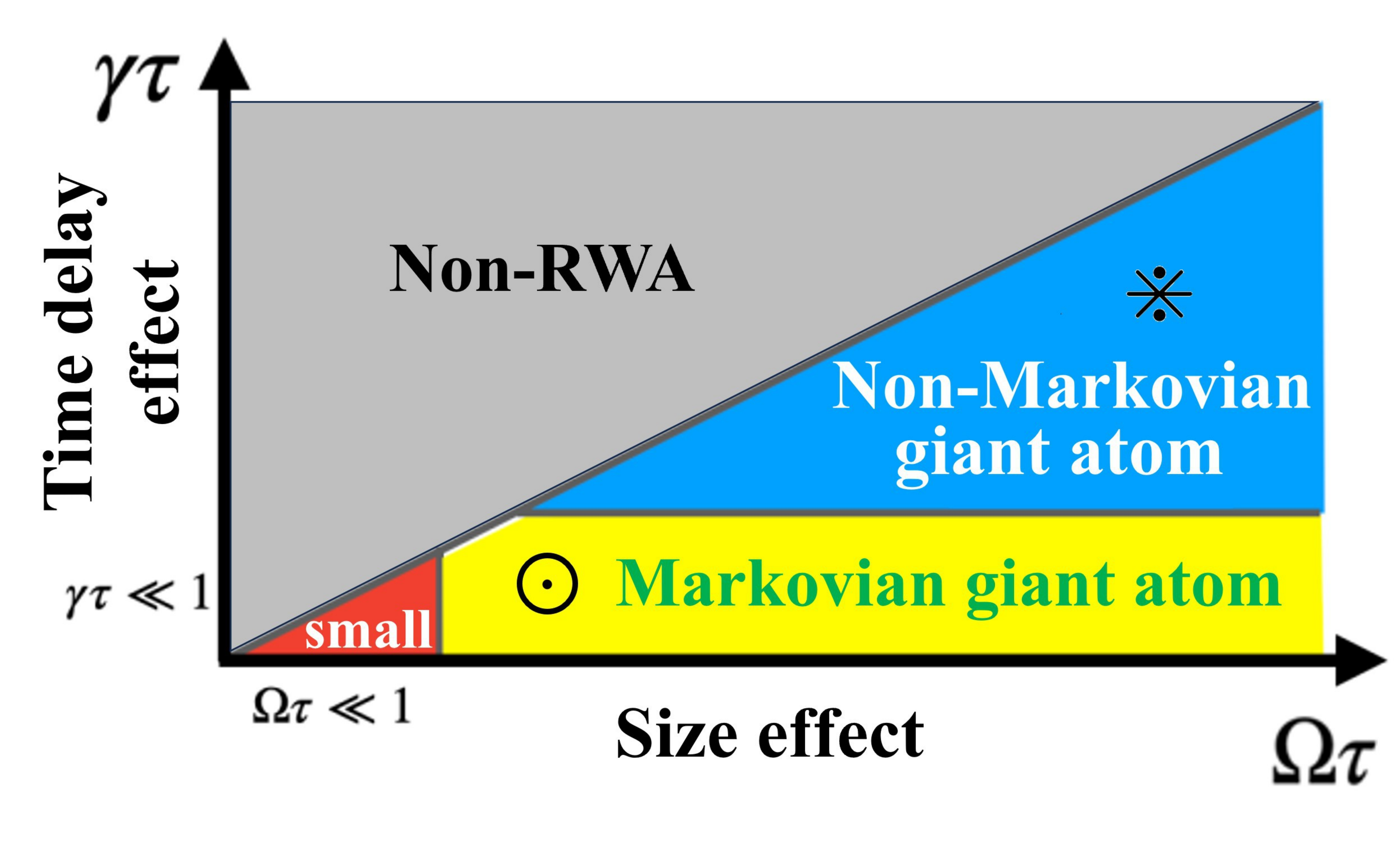}
\caption{Parameter space of giant atom spanned by the dimensionless parameters $\Omega\tau$ (size effect) and $\gamma\tau$ (time delay effect) with  $\Omega$ the atomic transition frequency, $\tau$ the travel time between two neighboring coupling points, and $\gamma$ the relaxation rate of single coupling point.}
\label{parameter-space}
\end{figure}

\section{Model Hamiltonian} \label{Model Hamitonian}
We investigate the general model of multiple two-level giant atoms coupled to a one-dimensional (1D) waveguide at multiple coupling points, with the total Hamiltonian described by 
\begin{eqnarray}\label{eq-H}
       H &=& \sum_{i}\hbar\Omega^i\sigma^i_{+}\sigma^i_{-}+\int^{+\infty}_{-\infty} \mathrm{d}k \hbar\omega_{k}\hat{a}_{k}^{\dagger} \hat{a}_{k} \\
       &&+ \sum_{m=1}^{N} \int^{+\infty}_{-\infty} g_{m}(e^{ik{x_{m}}} \hat{a}_{k} \sigma^i_{+} +{\rm H.c.}) \sqrt{|k|} \mathrm{d}k. \nonumber
\end{eqnarray}
The first term on the right-hand side (RHS) denotes the Hamiltonian of multiple giant atoms with $\sigma^i_{+} \equiv \ket{e_i}\bra{g_i}$ ($\sigma^i_- \equiv \ket{g_i}\bra{e_i}$) the raising (lowering) operator of the $i$-th atom, where $\ket{g_i}$ ($\ket{e_i}$) is the ground (excited) state with the atomic transition frequency $\Omega^i$. The second term on the RHS denotes the Hamiltonian of the 1D waveguide, where parameters $k$, $v$, and $\omega_k = \lvert k \rvert v$ are the wave vectors, velocities, and frequencies of the bosonic modes in the waveguide with the field operators $\hat{a}_k$ satisfying $[\hat{a}_k , \hat{a}^{\dag}_{k^{'}}]$ = $\delta(k-k{'})$. The third term on the RHS denotes the Jaynes-Cummings (JC) coupling Hamiltonian between giant atoms and bosonic modes, where the rotating-wave approximation(RWA) has been applied to derive the interaction term from experimental setups (see more details in Appendix ~\ref{one giant atom FORMULA}). 
We assume there are $N$ coupling points connecting the giant atoms to the waveguide. We label the coupling points from the leftmost to the rightmost with the coordinates $x_m$ ($m=1,2,\cdots,N$). The traveling times for bosonic modes between two neighboring coupling points are given by $\tau_{mm'} = |x_{m}-x_{m'}|/v$ $(m, m' = 1,2,...,N)$. The parameter $g_m$ is the coupling strength at the coupling point $x_m$ measured as an energy density over the wave-vector space. In Fig.~\ref{fig-model-2}(a), we illustrate the model system of one giant atoms with $N=3$ coupling points, and in Fig.~\ref{fig-model-2}(b), we illustrate the model system of two giant atoms with $N=4$ coupling points in total.

To proceed, we first need to specify the parameter space of giant atoms. We assume the dominant physics of giant atoms exchanging energy with the waveguide happens around the frequency $\Omega^i\approx \Omega$, the characteristic distance between neighboring coupling points is $L=v\tau$, and the decay rate of atoms via an individual coupling point is $\gamma$. The parameter space is spanned by two dimensionless parameters $\Omega L/v=\Omega\tau$ and $\gamma\tau$ that describe the interference size effects and the time delay effects, respectively. 
As shown in Fig.~\ref{parameter-space}, the parameter space can be divided into several regions. The red and yellow colored regions correspond to the Markovian regimes, where the atomic lifetime is much longer than the traveling time between neighboring coupling points ($\gamma\tau\ll 1$).
The red colored region corresponds to the long-wavelength limit ($\Omega\tau=\Omega L/v\ll 1$) of small atom regime, which is the standard local interaction assumption in the conventional quantum optics. The yellow colored region goes beyond the local interaction paradigm with considering the giant size effect of Markovian giant atoms. In contrast, the blue colored region is the regime of non-Markovian giant atom where both the size effects and the time-delay effects play roles in the dynamics. The gray parameter region goes beyond the RWA which is not studied in this work.

In many theoretical models, the coupling strengths at different coupling points $g_m$ and the position of coupling points $x_m$ are assumed to be fixed with desired values. However, in the realistic experimental situations, there are always some level of disorder both in the coupling strength and in the coupling positions. The main goal of this work is to investigate the effects of the two disorders on the giant atoms.

\section{Dark states with disorders}\label{sec-BIC}

\subsection{Equation of motion} \label{equations of motion and its results}
We first investigate the effects of coupling disorders on the dark states of giant atoms. As shown in Fig.~\ref{fig-model-2}(a), we consider a single giant atom with $N$ coupling points to the waveguide. The corresponding Hamiltonian is described by Eq.~(\ref{eq-H}) via taking $\Omega^1 = \Omega$ and $\sigma^1_{+(-)}=\sigma_{+(-)}$.
We study the process of spontaneous emission of the giant atom decaying into the waveguide. We assume the atom is initially in the excited state $\ket{e}$ and the field in the waveguide is in the vacuum state $\ket{vac}$. As the total number of atomic and field excitations is conserved in Eq.~(\ref{eq-H}) due to the JC interaction, we restrict the total state in the single-excitation subspace described by
\begin{eqnarray} \label{total-system-state}
    \ket{\Psi(t)}=\beta(t) \ket{e,vac} + \int \mathrm{d}k \alpha_{k}(t) \hat{a}_{k}^{\dagger} \ket{g,vac}. \ \ \ 
\end{eqnarray}
Here, the $\beta(t)$ describes the excited probability amplitude of giant atom and the integral describes the state of a single bosonic mode propagating in the waveguide. 
From Schr$\rm \ddot{o}$dinger equation $i \hbar \partial_t \ket{\Psi(t)} = H \ket{\Psi(t)}$, we derive the equation of motion (EOM) for the probability amplitude of the giant atom (see more details in Appendix~\ref{equations of motion of single giant atom}):
\bea \label{formal d_beta(t)}
   && \dfrac{\mathrm{d}}{\mathrm{d}t}\beta(t)=-i\Omega\beta(t)  \notag \\
   &&-\dfrac{\gamma}{2}\sum^N_{m,m^{'}=1}G_{m}G_{m^{'}}\beta(t-|\tau_{mm'}|)\Theta(t-|\tau_{mm'}|) \notag \\
   &&-i\sqrt{\dfrac{\gamma v}{4\pi}}\sum^{N}_{m=1}\int^{+\infty}_{-\infty}G_{m}e^{i(kx_{m}-\omega_{k}t)}\alpha_{k}(0)\mathrm{d}k
\eea
with $\Theta(x)$ the Heaviside step function. To derive Eq.~(\ref{formal d_beta(t)}), we have assigned the coupling strength $g_m=G_mg_0$ with $G_m$ a factor that can vary at different coupling points. The parameter $\gamma\approx 4\pi g_0^2\Omega/\hbar^2v^2$~\cite{guo2020prr} is the relaxation rate at single coupling point that is approximated as a constant over the relevant frequency range in the spirit of Weisskopf-Wigner theory~\cite{Scully_Zubairy_1997}.

For the spontaneous emission from the giant atom into the waveguide, we set $\beta(0)=1$ for the atomic initial state and $\alpha_{k}(0) = 0$ for the initial vacuum state of the bosonic field in the waveguide. The solution of $\beta(t)$ of Eq.~(\ref{formal d_beta(t)}) can be obtained by the Laplace transformation (see more details in Appendix~\ref{equations of motion of single giant atom}),
\begin{eqnarray} \label{eq-laplace}
    \beta(t)=&\sum\limits_n \dfrac{e^{{s_n}t}}{1-\dfrac{\gamma}{2}\sum^N\limits_{m,m^{'}=1}G_{m}G_{m^{'}}|\tau_{mm'}|e^{-|\tau_{mm'}|s_n} } . 
\end{eqnarray}
Here, the index $n$ denotes all kinds of modes propagating in the waveguide, which have complex frequencies  $s_n = -\kappa_n - i\Omega_n$ with the positive real number $\kappa_n$ denoting the decay exponent.
According to Eq.~(\ref{eq-laplace}), the probability of atom $|\beta(t)|^2$ can be expressed alternatively as 
\begin{eqnarray} \label{beta(t)_2}
    |\beta(t)|^2 = \left|\sum_n A_n e^{-\kappa_n t - i \Omega_n t} \right|^2 ,
\end{eqnarray}
where the coefficients $A_n$ can be determined from the RHS of Eq.~(\ref{eq-laplace}) using the residual theorem (see more details in Appendix~\ref{equations of motion of single giant atom}). In the long-time evolution limit, only the mode with the smallest $\kappa_{min}$ can survive, and thus the atomic probability becomes
\begin{equation}
    |\beta(t)|^2 \to \left|A_{min} \right|^2 e^{-2\kappa_{min}t}.
\end{equation}

\begin{figure}
\centering
\includegraphics[width=\linewidth]{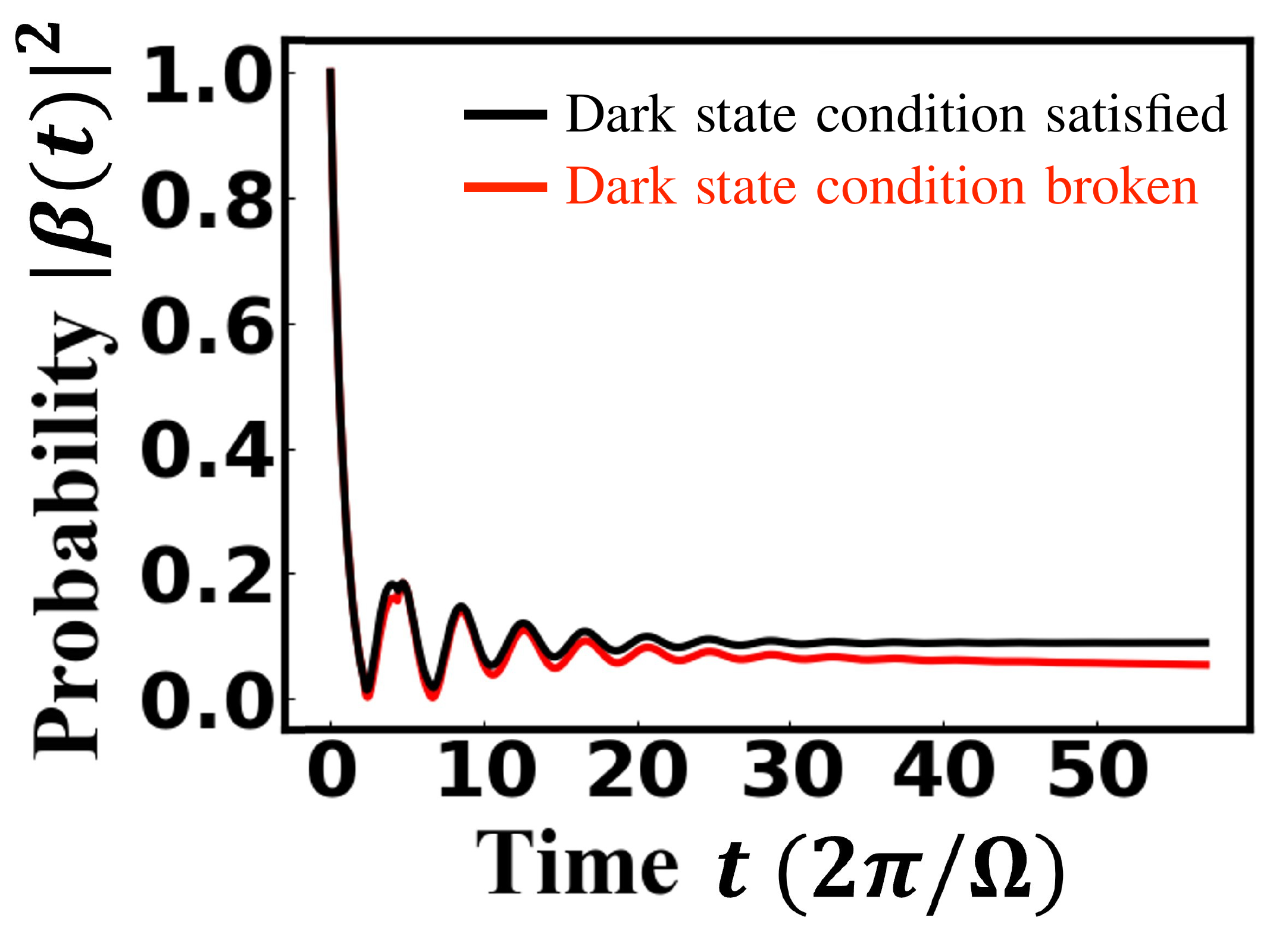}
\caption{The excitation probability of a typical non-Markovian giant atom as a function of evolution time. The atomic probability saturates at some finite values in the long-time limit when the dark state condition (\ref{eq-DS}) is satisfied (black line) with parameters: $\Omega\tau/2\pi=2.22$ and $\gamma\tau/2\pi=0.13$, but continues to decay when the condition (\ref{eq-DS}) is broken by the imposed disorders of coupling positions and strengths (red line) with parameters: $\Omega\tau_1/2\pi=2.231,\ \Omega\tau_2/2\pi=2.184$, $\gamma\tau_1/2\pi=0.1299,\ \gamma\tau_2/2\pi=0.1286$ and $\gamma\tau_3/2\pi=0.1329$. 
}
\label{non-Markovian-example}
\end{figure}

\begin{figure*}[htbp]
\centering
\includegraphics[width=1.0\textwidth]{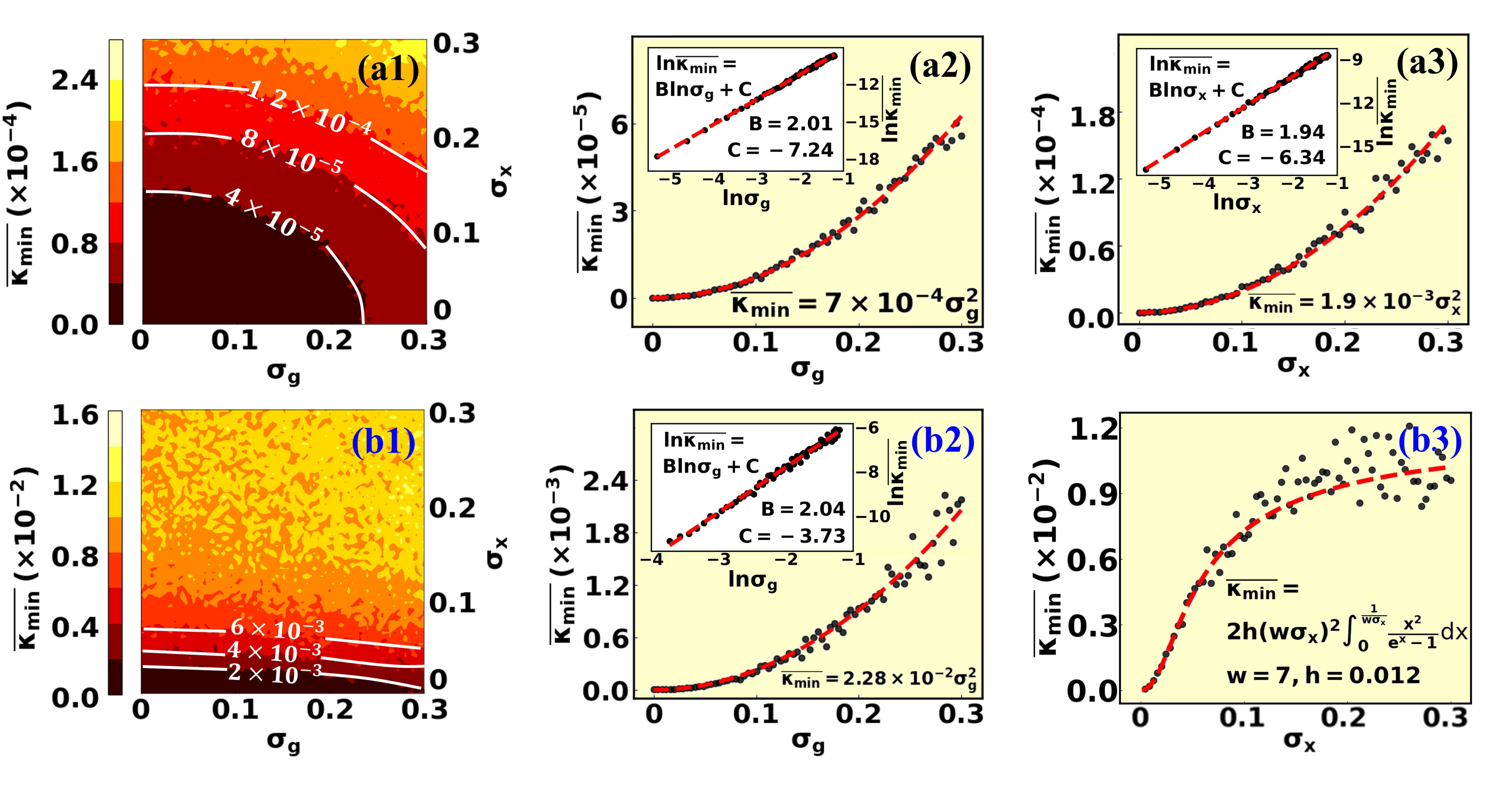}
\caption{Disorder effects on the dark state and BIC. {\bf (a1)} Averaged decay rate of giant atom $\overline{\kappa_{min}}$, cf. Eq.~(\ref{eq-kappamin}), as functions of the coupling-strength disorder deviation $\sigma_g$ and the coupling-position disorder deviation $\sigma_x$. {\bf (a2)} Averaged decay rate $\overline{\kappa_{min}}$ as a function of the coupling-strength disorder deviation $\sigma_g$ for the zero position disorder deviation $\sigma_x = 0$ with the log-log plot shown in the insert. {\bf (a3)} Averaged decay rate $\overline{\kappa_{min}}$ as a function of the coupling-position disorder deviation $\sigma_x$ for the zero coupling-strength disorder deviation $\sigma_g = 0$ with the log-log plot shown in the insert. For the plots in (a1)-(a3), parameters are chosen as $\gamma \tau / 2 \pi = 1.59 \times 10^{-4},\ \Omega \tau / 2 \pi = 0.33$ and the average of the decay rate is calculated over $200$ disorder samples. {\bf (b1)-(b3)} Same plots as shown in (a1)-(a3) for the parameter setting $\gamma \tau / 2 \pi = 0.13,\ \Omega \tau / 2 \pi = 2.22$ and the average over $100$ disorder samples. In the insert of {\bf(b3)}, $\sigma_x$ is extended to 0.4.}
\label{fig-Disorders-NM}
\end{figure*}

In general, all the modes will decay ultimately. However, when the system satisfies the dark state condition, i.e.,
\bea\label{eq-DS}
\Omega \tau = \dfrac{2n\pi}{N} - \dfrac{1}{2} N \gamma \tau \cot{\dfrac{n\pi}{N}} , \quad n \in \mathbb{Z}. 
\eea
There exists one particular mode (or more) with an exact zero decay exponent 
that does not decay despite the dissipative environment~\cite{guo2020prr}, as indicated by the black line in Fig.~\ref{non-Markovian-example}. Particularly, in the non-Markovian giant atom regime where the traveling time between coupling points is comparable to the lifetime of the giant atom, there will be the BICs of the bosonic field trapped by the coupling point in the waveguide. The time evolution of the bosonic field function  $\Phi(x,t) \equiv \dfrac{1}{\sqrt{2 \pi}} \int^{+\infty}_{-\infty} e^{ikx} \alpha_{k}(t) \mathrm{d}k$ is given by (see Appendix~\ref{equations of motion of single giant atom})
\bea \label{bosonic field}
\Phi(x,t)&=& \dfrac{1}{\sqrt{2 \pi}} \int^{+\infty}_{-\infty} e^{i(kx-\omega_{k}t)} \alpha_{k}(0) \mathrm{d}k \notag \\
&& - i \sqrt{\dfrac{\gamma}{2v}} \sum^{N}_{m=1} G_{m} \beta \left( t - \frac{|x - x_m|}{v} \right) \notag \\
&& \times\Theta \left( t - \frac{|x - x_m|}{v} \right).
\eea
Note that the BICs do not appear in the Markovian giant atom regime~\cite{guo2020prr}.

\subsection{Disorder effects}\label{sec-disorder}

To investigate the effects of disorders on the dark states and thus the BICs, we introduce some randomness to the coupling strengths and positions as follows.  We set the coupling strength $g_m = G_m g_0$ by adding Gaussian random numbers $\{G_m\}$ around unit ($\overline{G_m}=1$) with standard deviation $\sigma_g=\sqrt{\overline{(G_m-1)^2}}$. The level of coupling disorder can be tuned by the standard deviation $\sigma_g$. Similarly, by generating another set of Gaussian random number $\{L_m\}$ around zero ($\overline{L_m}=0$) with standard deviation $\sigma_x=\sqrt{\overline{L_m^2}}$, we set the coordinates of coupling points as $x_m=(m-1+L_m)v\tau$, where $m=1,2,\cdots,N$. Again, we can control the level of disorder of coupling positions by the standard deviation $\sigma_x$. Thus, the traveling time for bosonic modes between two neighboring coupling points $\tau_{mm'} = {|x_{m}-x_{m'}|}/{v}$ is also a random number. 
To obtain the decay rate, we first simulate the dynamics of atomic probability $|\beta(t)|^2$ for one disorder configuration as shown by the red curve in Fig.~\ref{non-Markovian-example}. Then, we take two time moments $t_1$ and $t_2$ in the long time limit from the dynamics of atomic probability and extract the minimum decay rate by 
\begin{equation}\label{eq-kappamin}
     \kappa_{min} = -\dfrac{1}{2(t_2-t_1)} \ln \left(\dfrac{|\beta(t_2)|^2}{|\beta(t_1)|^2} \right).
\end{equation}
Finally, we average the extracted decay rate $\overline{\kappa_{min}}$ over a sufficient number of disorder configurations.

In Fig.~\ref{fig-Disorders-NM}(a1), we plot the decay exponent $\overline{\kappa_{min}}$ averaged $200$ random configurations as functions of the standard deviations of coupling-strength disorder and the coupling-position disorder for a giant atom with $N=3$ coupling points in the Markovian regime (as marked by the symbol $\odot$ in Fig.~\ref{parameter-space}). As expected, the extracted decay rate increases monotonically as both the two disorder deviations $\sigma_g$ and $\sigma_x$ increase.
To further explore the explicit dependence of the decay rate on each disorder, we plot the decay rate as a function of the coupling-strength disorder deviation $\sigma_g$ with zero coupling-position disorder deviation $\sigma_x=0$ in Fig.~\ref{fig-Disorders-NM} (a2).  The black points correspond to the original data from numerical calculations, and the red dashed line is the fitting line of these points with the quadratic function $\overline{\kappa_{min}}\propto \sigma_g^2$. In the insert, we also plot the logarithm of the averaged decay rate $\ln\overline{\kappa_{min}}$ as a function of the logarithm of the disorder deviation $\ln\sigma_g$, where the extracted slope of the linear relationship confirms the quadratic scaling behavior. Similarly, in Fig.~\ref{fig-Disorders-NM} (a3), we plot the decay rate as a function of the coupling-position disorder deviation $\sigma_x$ while keeping the coupling-strength disorder deviation zero $\sigma_g=0$, which also shows a quadratic scaling behavior (see the fitting and the insert).

In Fig.~\ref{fig-Disorders-NM}(b1), we plot the decay exponent $\overline{\kappa_{min}}$ averaged over $100$ random configurations as functions of the two disorder standard deviations with $N=3$ coupling points in the non-Markovian regime (as marked by the symbol $\divideontimes$ in Fig.~\ref{parameter-space}). It shows that the decay exponent is further suppressed by the non-Markovian effect. In Fig.~\ref{fig-Disorders-NM} (b2), we plot the decay rate as a function of the coupling-strength disorder deviation $\sigma_g$ with zero coupling-position disorder deviation $\sigma_x=0$. Similar to the Markovian case, the decay rate still shows a quadratic scaling behavior as a function of coupling-strength disorder $\overline{\kappa_{min}}\propto \sigma_g^2$, cf. the fitting and the insert in Fig.~\ref{fig-Disorders-NM} (b2). However, the averaged decay rate $\overline{\kappa_{min}}$ exhibits a very different behavior as a function of the coupling-position disorder $\sigma_x$, where the decay rate saturates as $\sigma_x$ increases, as shown by Fig.~\ref{fig-Disorders-NM} (b3). 
In fact, we find the numerical data can be fit by the extended Debye function~\cite{Debye-function} of order two defined via
\begin{eqnarray} \label{extended-debye-function}
     \overline{\kappa_{min}} &=& \mathcal{D}_2(\sigma_x;h,w) \nn\\
     &\equiv& 2h(w\sigma_x)^2\int^{\frac{1}{w \sigma_x}} _0 \dfrac{x^{2}}{e^x - 1} \mathrm{d}x , \\
     &=& 
     \left \{ \begin{array}{lll}
    4h w^{2} \zeta(3) \sigma_{x}^{2} & w\sigma_x \ll 1,\\
    h \left( 1- \dfrac{1}{3 w \sigma_x} + \dfrac{1}{24 w^{2} \sigma_{x}^{2}} -  \cdots \right) & w\sigma_x \gg 1.
 \end{array} \right. \notag
\end{eqnarray}
Here, the parameter $w$ denotes the width of this curve, $h$ denotes the height of this curve, $\zeta(x)$ is the Riemann zeta function, and $\zeta(3)$ is known as Ap$\mathrm{\Acute{e}}$ry's constant. We can use two parameters ($w$ and $h$) to fit the scatter points, cf. the red curve in Fig.~\ref{fig-Disorders-NM} (b3). The extended Debye function recovers the quadratic scaling behavior $\overline{\kappa_{min}}\propto \sigma^2_x$ when the coupling-position disorder approaches zero $\sigma_x\rightarrow 0$, but becomes a constant $\overline{\kappa_{min}}=h$ for the large coupling-position disorder.
In the limit of $\sigma_x\rightarrow \infty$, each coupling point is allowed to be located randomly on the waveguide everywhere with equal distribution. Thus, the giant atom explores all the possible configurations coupled to the waveguide with the effective decay rate ranging from zero to $N^2\gamma$~\cite{Kockum2014pra}.  The final plateau of the decay rate should be some averaged value smaller than $N^2\gamma$. A further analytical framework is needed to calculate the accurate value of the plateau.

\begin{figure}
\centering
\includegraphics[width=0.48\textwidth]{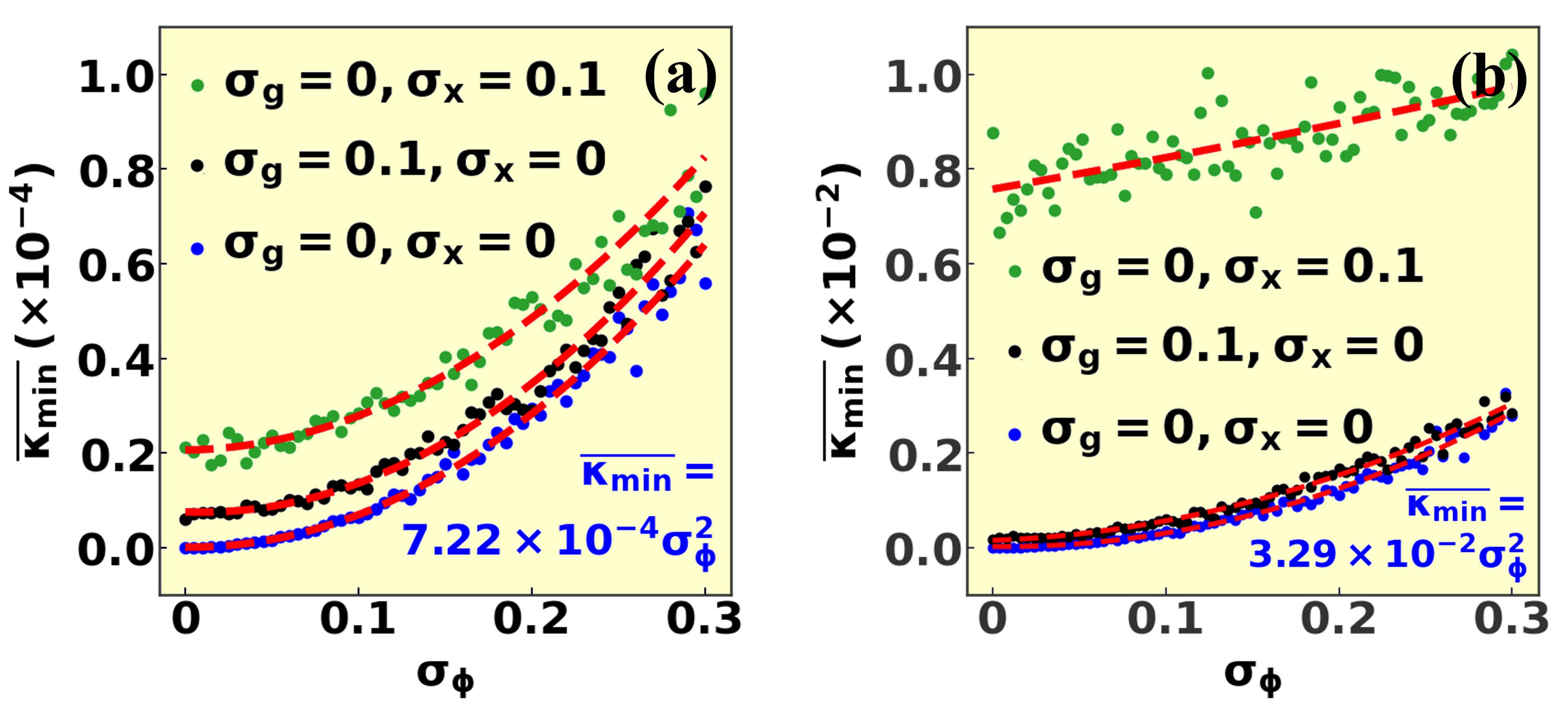}
\caption{Disorder effects of the coupling phase. {\bf (a)} Averaged decay rate $\overline{\kappa_{min}}$ as a function of the coupling-phase disorder $\sigma_{\phi}$ for different coupling and position disorder settings (indicated by different colors) in the Markovian regime with parameters: $\gamma \tau / 2 \pi = 1.59 \times 10^{-4},\ \Omega \tau / 2 \pi = 0.33$.  {\bf (b)} Plot of $\overline{\kappa_{min}}$ as a function of $\sigma_{\phi}$ in the non-Markovian regime with the parameters: $\gamma \tau / 2 \pi = 0.13,\ \Omega \tau / 2 \pi = 2.22$. In both figures, the data are calculated over $100$ disorder samples.}
\label{fig-phi}
\end{figure}

\subsection{Effects of coupling-phase disorder}

Above, we have modeled the coupling strengths $g_m$ as real scalars. In the experiments, however, the coupling strength could acquire a complex phase due to impedance fluctuations. To investigate the role of fluctuating phases, we set $g_m = G_m g_0 e^{i \phi_m}$ by adding Gaussian random numbers $\{\phi_m\}$ with zero mean value ($\overline{\phi_m}=0$) and the standard deviation $\sigma_{\phi}=\sqrt{\overline{\phi_m^2}}$.
Thus, Eq.~(\ref{formal d_beta(t)}) is modified as follows
\bea
   \dfrac{\mathrm{d}\beta(t)}{\mathrm{d}t}&=&-i\Omega\beta(t) -\dfrac{\gamma}{2}\sum^N_{m,m^{'}=1}G_{m}G_{m^{'}} e^{i(\phi_m - \phi_{m^{'}})}\nn\\
   &&\times\beta(t-|\tau_{mm'}|)\Theta(t-|\tau_{mm'}|).
\eea
In Fig.~\ref{fig-phi} (a), we plot the averaged decay rate $\overline{\kappa_{min}}$ as a function of the complex phase disorder $\sigma_{\phi}$ for different settings of 
coupling and position disorders in the Markovian regime. The data of the blue points represents the result with zero coupling and position disorders ($\sigma_g=0$, $\sigma_x=0$). The black and green points show the results for disorder settings of ($\sigma_g = 0.1$, $\sigma_x = 0$) and ($\sigma_g = 0$, $\sigma_x = 0.1$), respectively. 
In Fig.~\ref{fig-phi} (b), we plot the averaged decay rate $\overline{\kappa_{min}}$ as a function of the complex phase of coupling-strength disorder deviation $\sigma_{\phi}$ in the non-Markovian regime for different coupling and position disorder choices. The green points ($\sigma_g = 0.1$, $\sigma_x = 0$) with higher values mean the decay rate is more sensitive to the position disorder than the coupling disorder  ($\sigma_g = 0$, $\sigma_x = 0.1$).
As expected, in both regimes, the averaged decay rate increases monotonically as $\sigma_{\phi}$ increases, following a quadratic scaling behaviour.
Therefore, the predicted phenomena of the dark state and BIC are robust in the presence of coupling phase disorders.

\section{Decoherence-free interaction with disorder}\label{sec-DFI}

\begin{figure*}
\centering
\includegraphics[width=0.9\textwidth]{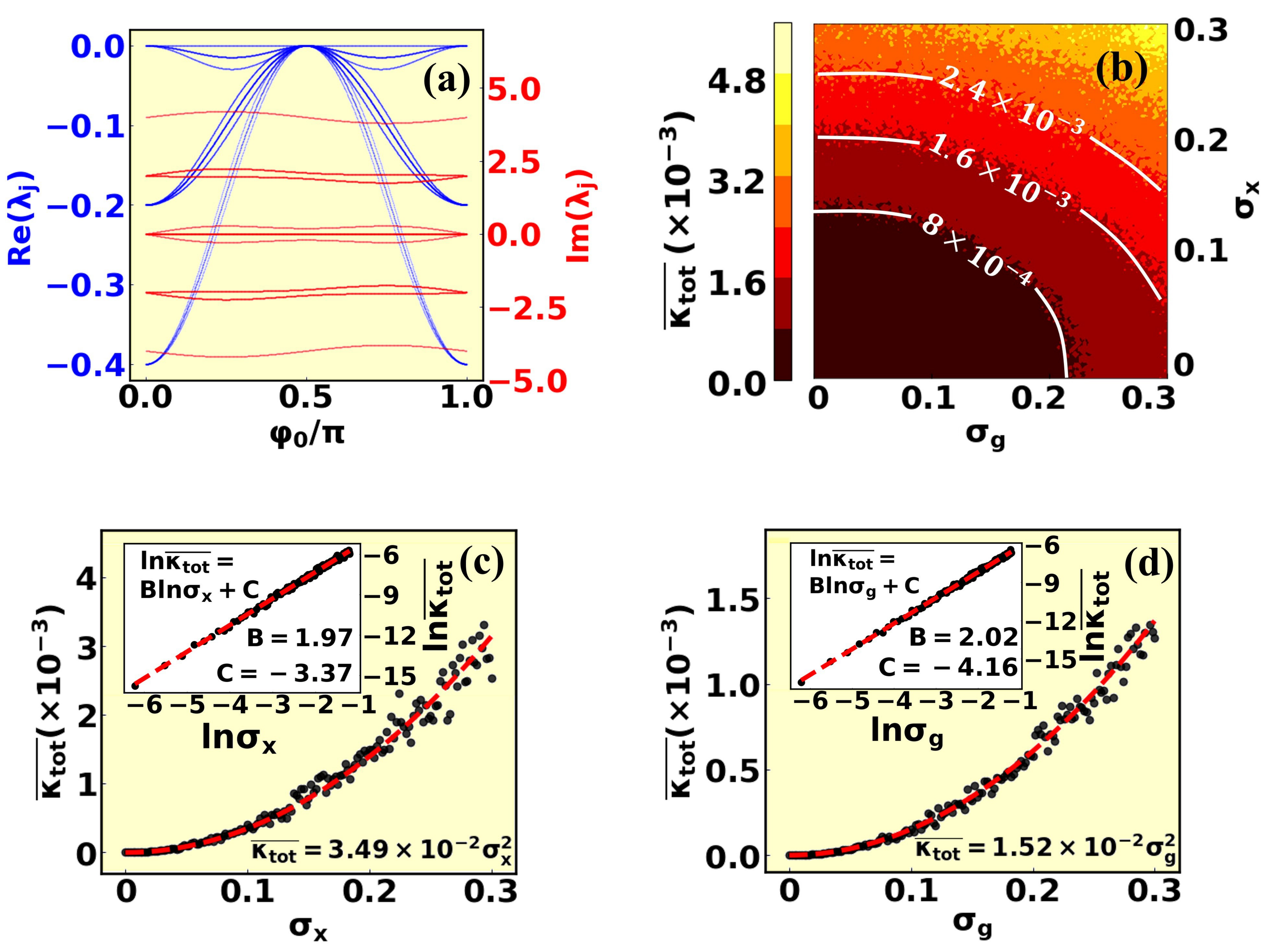}
\caption{Disorder effects on DFI. {\bf (a)} Real part (blue line) and imaginary part (red line) of eigenvalues of 
the vectorized master equation, cf. Eq.~(\ref{eq-L}), as functions of the propagating
 phase $\varphi_0$ between nearest coupling points. {\bf (b)} Averaged total decay rate of braided giant atoms $\overline{\kappa_{tot}}$, cf. Eq.~(\ref{eq-kappatot}),  as functions of the coupling-strength disorder deviation $\sigma_g$ and the coupling-position disorder deviation $\sigma_x$ for the parameter setting $\gamma = 4.78\times 10^{-4}, \varphi_0 = \pi /2$. The average is calculated over $100$ disorder samples. {\bf (c)} Averaged total decay rate $\overline{\kappa_{tot}}$ as a function of the coupling-position disorder deviation $\sigma_x$ for the zero coupling-strength disorder deviation $\sigma_g = 0$ with the log-log plot shown in the insert.  {\bf (d)} Averaged total decay rate $\overline{\kappa_{tot}}$ as a function of the coupling-strength disorder deviation $\sigma_g$ for the zero coupling-position disorder deviation $\sigma_x = 0$ with the log-log plot shown in the insert. }
\label{fig-DFI-2}
\end{figure*}

We now investigate another important phenomenon related to giant-atom physics, i.e., the DFI between multiple giant atoms~\cite{Kockum2018prl}, which appears in the Markovian regime where the traveling time between coupling points can be neglected compared to the decay time of giant atoms. 
We consider the case of two giant atoms coupled to a 1D waveguide at two points as depicted in Fig.~\ref{fig-model-2}(b). The two giant atoms have three topologies for the configuration of the coupling points, i.e., the separated, the braided, and the nested giant atoms. The case of braided giant atoms is particularly interesting as such a coupling configuration supports the DFI.

Using the SLH formalism \cite{Gough_2008,Gough2007TheSP,Combes_2017} for cascaded quantum systems \cite{Carmichael:1993xyz,Gardiner_1993,zoller1997}, the master equation for multiple braided giant atoms can be written as ~\cite{Kockum2018prl},
\bea \label{master-equation}
    \dot{\rho} &= & -i \left [ \omega^{'}_a \dfrac{\sigma_{z}^{a}}{2} +\omega^{'}_b \dfrac{\sigma_{z}^{b}}{2} + f(\sigma_{-}^{a} \sigma_{+}^{b} + \sigma_{+}^{a} \sigma_{-}^{b}) , \rho \right ] \notag \\
    &&+ \Gamma_a \mathcal{D} [\sigma_{-}^{a}] \rho + \Gamma_b \mathcal{D} [\sigma_{-}^{b}] \rho \\
    &&+ \Gamma_{\rm{coll}} \left[ \left( \sigma_{-}^{a} \rho \sigma_{+}^{b} - \dfrac{1}{2} \{ \sigma_{+}^{a} \sigma_{-}^{b} , \rho \}\right) + \rm{H.c.} \right], \notag  
\eea 
where $\mathcal{D} [X] \rho = X \rho X^{\dagger} - \dfrac{1}{2} X^{\dagger} X \rho - \dfrac{1}{2} \rho X^{\dagger} X$ is the Lindblad operator. The normalized atomic transition frequencies are given by
\bea
 \left \{ \begin{array}{lll}
\omega'_{a}&=&\omega_{a} + \sqrt{\gamma_{1} \gamma_{3}} \sin(\varphi_{1} + \varphi_{2}),\\
 \omega'_{b}&=&\omega_{b} + \sqrt{\gamma_{2} \gamma_{4}} \sin(\varphi_{2} + \varphi_{3}),
 \end{array} \right.
\eea
where $\omega_{a(b)}$ is the normalized transition frequency of atoms, cf. the bare atomic frequency $\Omega^{a(b)}$ in Eq.~(\ref{eq-H}), parameter $\varphi_m \equiv \omega_a \lvert x_{m+1} - x_{m} \rvert/v$ depending on the coupling positions is the propagating phases, and $\gamma_m\approx 4\pi g_m^2\Omega/\hbar^2v^2$ is the decay rate of individual coupling points. The parameter $f$ is the strength of the exchange interaction between the atoms in the form of
\bea\label{eq-f}
f&=&\dfrac{1}{2} [ \sqrt{\gamma_1 \gamma_2} \sin\varphi_1 + \sqrt{\gamma_2 \gamma_3} \sin\varphi_2  \\
    &&+ \sqrt{\gamma_3 \gamma_4} \sin\varphi_3 + \sqrt{\gamma_1 \gamma_4} \sin(\varphi_1 + \varphi_2 + \varphi_3)] \notag.
\eea
The parameter $\Gamma_{a(b)}$ is the individual atomic relaxation rate to the waveguide given by
\bea
 \left \{ \begin{array}{lll}
\Gamma_{a}&=& \gamma_1 + \gamma_3 + 2\sqrt{\gamma_1 \gamma_3} \cos(\varphi_1 + \varphi_2) ,\\
 \Gamma_{b}&=& \gamma_2 + \gamma_4 + 2\sqrt{\gamma_2 \gamma_4} \cos(\varphi_2 + \varphi_3).
 \end{array} \right.
\eea
The parameter $\Gamma_{\rm{coll}}$ is the collective relaxation rate for the atoms given by
\bea
\Gamma_{\rm{coll}}&=& \sqrt{\gamma_1 \gamma_2} \cos\varphi_1 + \sqrt{\gamma_2 \gamma_3} \cos\varphi_2 \\
    &&+ \sqrt{\gamma_3 \gamma_4} \cos\varphi_3 + \sqrt{\gamma_1 \gamma_4} \cos(\varphi_1 + \varphi_2 + \varphi_3).\nn
\eea
More details for the derivation of the master equation are provided in Ref.~\cite{Kockum2018prl}.

In order to deal with disorders numerically, we vectorize the density matrix and rewrite the master equation in the form of 
\begin{equation}\label{eq-L}
\dfrac{\mathrm{d}\vec{\rho}}{\mathrm{d}t} = \mathcal{L} \vec{\rho},
\end{equation}
see more technical details and the explicit expression of superoperator $\mathcal{L}$ in Appendix~\ref{Derive the vectorized density matrix}. Solving the eigen problem of Eq.~(\ref{eq-L}) in the basis of two atoms, we can have in total sixteen complex eigenvalues together with their eigenvectors. Given the eigenvector $\vec{v}_j$ and the eigenvalue $\lambda_j$ ($j = 1, 2, 3,..., 16$), the solution of master equation in form of the vectorized density matrix is 
\bea
    \vec{\rho}(t) = \sum_j c_j e^{\lambda_j t} \vec{v}_j.
\eea
Here, the coefficients $c_j$ are determined by the initial state of the density matrix. The real part of complex eigenvalues represents the decay rate, and the imaginary parts denote the frequency. In particular, the real part of the eigenvalue should not be positive, as the giant atoms always dissipate energy when interacting with the waveguide. 
In Fig.~\ref{fig-DFI-2}(a), we plot the real and imaginary parts of all the eigenvalues as functions of propagating phase $\varphi_0$ in the ideal case of equally distributed coupling points without disorders. When the propagating phase takes the value of $\varphi_0 = {\pi}/{2}$ and all the individual decay rates are identical $\gamma_m=\gamma$, the real parts of all the eigenvalues become exactly zero. In this case, the two atoms do not decay into the waveguide but still have finite interaction $f=\gamma$, cf. Eq.~(\ref{eq-f}), which is the essence of the DFI.

However, the disorders of coupling strengths and positions will violate the ideal DFI condition and result in the decay of giant atoms into the waveguide. We set the individual decay rate $\gamma_m= |G_m|^2\gamma$ and the propagating phase $\varphi_m=(1+L_m)\varphi_0$ at each coupling position, where $G_m$ and $L_m$ are Gaussian random numbers introduced in Sec.~\ref{sec-disorder}. To study the effects of disorders on the DFI, we summarize the real parts of all the eigenvalues, obtaining the total decay rate 
\begin{equation}\label{eq-kappatot}
    \kappa_{tot}\equiv - \sum_j \mathrm{Re}(\lambda_j).
\end{equation}
In Fig.~\ref{fig-DFI-2}(b), we average the extracted total decay rate $\overline{\kappa_{tot}}$ over sufficient number of disorder configurations and plot it as functions of the standard deviations of the coupling-strength disorder deviation $\sigma_{g}$ and the coupling-position disorder deviation $\sigma_{x}$.
We find that DFI is indeed suppressed by both disorders. The averaged total decay rate increases monotonically as both disorder increase.
To further investigate the effects of each disorder on the DFI, we plot the averaged total decay rate $\overline{\kappa_{tot}}$ as a function of one type of disorder standard deviation while keeping the other disorder standard deviation zero in Fig.~\ref{fig-DFI-2} (c) and (d). In the insert, we plot the logarithm of averaged total decay rate $\ln{\overline{\kappa_{tot}}}$ as a function of the logarithm of the disorder standard deviation $\ln\sigma_{x}$ or $\ln\sigma_{g}$ to reveal the scaling relationship between the averaged total decay rate and the disorders. We find the averaged total decay rate follows a polynomial behavior as a function of disorder, i.e., $\overline{\kappa_{tot}}\propto \sigma^\alpha_{x(g)}$ with the exponent $\alpha\approx 2$ extracted from the numerical fitting. In conclusion, the coupling position and strength disorders basically have similar suppression effects on the DFI in the Markovian regime for giant atoms.

\section{Summary}\label{sec-sumlook}

In summary, we have studied the system of giant atom(s) coupled to a 1D waveguide via multiple coupling points with considering the disorders of coupling positions and coupling strengths. We investigated the disorder effects on the dark states and BICs for the spontaneous emission of a single giant atom. We find that the decay rate increases monotonically as the standard deviations of the two disorders increase. In the Markovian regime, we have identified that the decay rates follow a quadratic scaling behavior as functions of both disorders. In the non-Markovian regime, the quadratic fitting is still valid for the coupling-strength disorder but breaks down for the coupling-position disorder, instead of an extended Debye function of order two, cf. Eq.~(\ref{extended-debye-function}). 

Our work provides an important insight into designing the giant-atom system in the experiments. The system of giant atoms is robust to the disorders of both coupling positions and coupling strengths in the Markovian regime, as both disorders suppress phenomena such as dark states and DFI in a quadratic way. However, to observe the non-Markovian phenomenon like (oscillating) BICs, it needs more precision to control the disorder of coupling positions than that of the coupling strengths. \\


\section*{Acknowledgement}
This work was supported by the National Natural Science Foundation of China (Grant No. 12475025).

\section*{Data Availability}
The data that support the findings of this article are openly available~\cite{han2025zendo}.

\bigskip

\onecolumngrid

\appendix

\section{Hamiltonian} \label{one giant atom FORMULA}
The transmon (giant atom) is coupled capacitively to the waveguide. The total Hamiltonian of the transmon coupled to a waveguide is given by \cite{Koch2007pra}
\bea \label{total H}
    H_{\rm tr} = \dfrac{(2e)^2}{2C_{\Sigma}} (\hat{n} - \hat{n}_s)^2 - E_J \cos{\hat{\varphi}} = 4 E_C \hat{n}^2 - E_J \cos{\hat{\varphi}} - 8 E_C \hat{n} \hat{n}_s + 4 E_C \hat{n}^2_s ,
\eea
where $\hat{n}_s$ is the offset charge of the transmon measured in units of the Cooper pair charge 2e, $E_J$ is the Josephson energy, and $E_C = \dfrac{e^2}{2C_{\Sigma}}$ is the charging energy with $C_{\Sigma}$ the total capacitance of the transmon. By defining the operators $\hat{b}$, $\hat{b}^{\dagger}$ via $\hat{\varphi} = \sqrt{\dfrac{\eta}{2}} (\hat{b} + \hat{b}^{\dagger})$, $\hat{n} = -i \sqrt{\dfrac{1}{2 \eta}} (\hat{b} - \hat{b}^{\dagger})$ with $\eta \equiv \sqrt{\dfrac{8 E_C}{E_J}}$, $[\hat{b} , \hat{b}^{\dagger}] = 1$, the free transmon Hamiltonian is,
\bea
    H_{\rm tr,free} =&& 4 E_C \hat{n}^2 - E_J \cos{\hat{\varphi}} = 4 E_C (-i \sqrt{\dfrac{1}{2 \eta}} (\hat{b} - \hat{b}^{\dagger}))^2 - E_J \cos{\sqrt{\dfrac{\eta}{2}} (\hat{b} + \hat{b}^{\dagger})} \notag \\
    =&& - 4 E_C \cdot \dfrac{1}{2 \eta} (\hat{b} - \hat{b}^{\dagger})^2 - E_J (1 - \dfrac{1}{2!} \cdot \dfrac{\eta}{2} (\hat{b} + \hat{b}^{\dagger})^2 + \dfrac{1}{4!} \cdot \dfrac{1}{4} \eta^2 (\hat{b} + \hat{b}^{\dagger})^4) \notag \\
    =&& \sqrt{\dfrac{E_J E_C}{2}} (2\hat{b}^{\dagger} \hat{b} + 2\hat{b} \hat{b}^{\dagger}) - \dfrac{1}{12} E_C (\hat{b} + \hat{b}^{\dagger})^4) - E_J \approx \hbar \omega_0 (\hat{b}^{\dagger} \hat{b} + \dfrac{1}{2}) - \chi (\hat{b} + \hat{b}^{\dagger})^4,
\eea
where $\omega_0 = \sqrt{8 E_J E_C} / \hbar$ is known as the Josephson plasma frequency and $\chi = \dfrac{E_C}{12}$ is the nonlinearity.

The electric potential ﬁeld $\hat{\phi}(x,t)$ in the waveguide can be described by \cite{Peropadre2013}
\begin{equation}
    \hat{\phi}(x,t) = -i \sqrt{\dfrac{\hbar Z_0 v}{4 \pi}} \int^{+\infty}_{-\infty} \mathrm{d} k \sqrt{\omega_k} (\hat{a}_k e^{-i (\omega_k t - kx)} - \rm H.c.),
\end{equation}
$\hat{a}_k$ is the annihilation operator of the waveguide mode with wave vector $k$, satisfying the commutation relations $[\hat{a}_k , \hat{a}^{\dagger}_{k'}] = \delta(k - k')$, $Z_0$ is the characteristic impedance of the waveguide, and $v$ is the velocity of SAWs or the speed of light
(microwaves) with the dispersion relation $\omega_k = \lvert k \rvert v$.

The coupling between the transmon and the waveguide is described by the term $H_{\rm int} = - 8 E_C \hat{n} \hat{n}_s$ in Eq.~(\ref{total H}) with the offset charge $\hat{n}_s = \dfrac{1}{2e} \sum_{m=1}^N C_m \hat{\phi}(x_m,t)$, where $C_m$ and $x_m$ are the effective capacitance and the position of each coupling point, respectively. Thus,  the interaction Hamiltonian is
\bea
    H_{\rm int} &=& - 8 E_C \hat{n} \hat{n}_s \notag \\
    &=& - 8 E_C \left[ -i \sqrt{\dfrac{1}{2 \eta}} (\hat{b} - \hat{b}^{\dagger}) \right] \dfrac{1}{2e} \sum_{m=1}^N C_m \left\{-i \sqrt{\dfrac{\hbar Z_0 v}{4 \pi}} \int^{+\infty}_{-\infty} \mathrm{d} k \sqrt{\omega_k} \left[ \hat{a}_k e^{-i (\omega_k t - kx_m)} - \rm H.c. \right] \right\} \notag \\
    &=& \dfrac{4 E_C}{e} \sqrt{\dfrac{1}{2 \eta}} \sqrt{\dfrac{\hbar Z_0 v}{4 \pi}} \sum_{m=1}^N C_m \int^{+\infty}_{-\infty} \mathrm{d} k \sqrt{\lvert k \rvert} \sqrt{v} \left[ \hat{a}_k e^{-i (\omega_k t - kx_m)} - {\rm H.c.} \right] (\hat{b} - \hat{b}^{\dagger}) \notag \\
    &=& - \sum_{m=1}^N g_m \int^{+\infty}_{-\infty} \mathrm{d} k \sqrt{\lvert k \rvert} \left[ \hat{a}_k e^{i kx_m} - \hat{a}_k^{\dagger} e^{-i kx_m} \right] (\hat{b}^{\dagger} - \hat{b}).
\eea
We define the coupling strength
\begin{equation}
    g_m = \dfrac{4 E_C}{e} v \sqrt{\dfrac{1}{2 \eta}} \sqrt{\dfrac{\hbar Z_0}{4 \pi}} C_m,
\end{equation}
and adopt the RWA by dropping counter-rotating terms like $\hat{a}_k^{\dagger} \hat{b}^{\dagger}$ and $\hat{a}_k \hat{b}$.
Thus, $H_{\rm int}$ can be written as,
\begin{equation}
    H_{\rm int} = \sum_{m=1}^N \int^{+\infty}_{-\infty} g_m (\hat{b}^{\dagger} \hat{a}_k e^{i kx_m} + {\rm H.c.}) \sqrt{\lvert k \rvert} \mathrm{d} k.
\end{equation}

The total Hamiltonian, including the waveguide, is
\begin{equation} \label{Hamiltonian with waveguide}
    H = \hbar \omega_0 (\hat{b}^{\dagger} \hat{b} + \dfrac{1}{2}) - \chi (\hat{b} + \hat{b}^{\dagger})^4 + \int^{+\infty}_{-\infty} \mathrm{d} k \hbar \omega_k \hat{a}_k^{\dagger} \hat{a}_k + \sum_{m=1}^N \int^{+\infty}_{-\infty} g_m (\hat{b}^{\dagger} \hat{a}_k e^{i kx_m} + {\rm H.c.}) \sqrt{\lvert k \rvert} \mathrm{d} k,
\end{equation}
using $[\hat{b} , \hat{b}^{\dagger}] = 1$, the second term of Eq.~(\ref{Hamiltonian with waveguide}) can be written as,
\bea \label{chi-b^dagger}
    - \chi (\hat{b} + \hat{b}^{\dagger})^4 &=& - \chi (\hat{b}^2 + \hat{b}^{\dagger} \hat{b} + \hat{b} \hat{b}^{\dagger} + \hat{b}^{\dagger 2})(\hat{b}^2 + \hat{b}^{\dagger} \hat{b} + \hat{b} \hat{b}^{\dagger} + \hat{b}^{\dagger 2}) \notag \\
    &=& - \chi (\hat{b}^4 + 6 \hat{b}^2 + 4 \hat{b}^{\dagger} \hat{b}^3 + 12 \hat{b}^{\dagger} \hat{b} + 6 \hat{b}^{\dagger 2} \hat{b}^2 + 6 \hat{b}^{\dagger 2} + 4 \hat{b}^{\dagger 3} \hat{b} + \hat{b}^{\dagger 4} + 3).
\eea
For the spontaneous emission process, the RWA guarantees that there is only one excitation either in the atomic state or in the waveguide. In this case, only the lowest two levels of the transmon, i.e., the ground state $\ket{g}$ and the ﬁrst excited state $\ket{e}$, are involved in the dynamics. Thus, only $\hat{b}^{\dagger} \hat{b}$ term can be remained in Eq.~(\ref{chi-b^dagger}). By defining the lowering operator $\sigma_- \equiv \ket{g} \bra{e}$ and raising operator $\sigma_+ \equiv \ket{e} \bra{g}$, we can write the Hamiltonian in the single-excitation subspace
\bea \label{Hamiltonian in the single-excitation subspace}
    H = \hbar \Omega \sigma_+ \sigma_- + \int^{+\infty}_{-\infty} \mathrm{d} k \hbar \omega_k \hat{a}_k^{\dagger} \hat{a}_k + \sum_{m=1}^N \int^{+\infty}_{-\infty} g_m (e^{i kx_m} \hat{a}_k \sigma_+ + {\rm H.c.}) \sqrt{\lvert k \rvert} \mathrm{d} k.
\eea
Here, the atomic transition frequency $\Omega = \omega_0 - 12 \chi / \hbar = \sqrt{8 E_C E_J} / \hbar - E_C / \hbar$ is the level spacing of the two lowest levels.

Moreover, expanding Eq.~(\ref{Hamiltonian in the single-excitation subspace}) into multiple($i$) two-level giant atoms coupled to a 1D waveguide at multiple($m$) coupling points, the total Hamiltonian can be described by
\begin{equation}
       H = \sum_{i}\hbar\Omega^i\sigma^i_{+}\sigma^i_{-}+\int^{+\infty}_{-\infty} \mathrm{d}k \hbar\omega_{k}\hat{a}_{k}^{\dagger} \hat{a}_{k} + \sum_{m=1}^{N} \int^{+\infty}_{-\infty} g_{m}(e^{ik{x_{m}}} \hat{a}_{k} \sigma^i_{+} +{\rm H.c.}) \sqrt{|k|} \mathrm{d}k,
\end{equation}
which is Eq.~(\ref{eq-H}) in the main text.

\section{Equations of motion of single giant atom} \label{equations of motion of single giant atom}
In this appendix, we derive the equation of motion for the probability amplitude  $\beta (t)$ of the single giant atom.
Substituting Eq.~(\ref{Hamiltonian in the single-excitation subspace}) and Eq.~(\ref{total-system-state}) into the Schroedinger equation $H\ket{\Psi(t)} = i\hbar \dfrac{\partial}{\partial t} \ket{\Psi(t)}$, we have
\bea \label{subsititute schodinger equation}
    H\ket{\Psi(t)} &=& \hbar\Omega \beta(t) \ket{e,vac} + \hbar \int \mathrm{d}k \omega_{k} \alpha_{k}(t) \hat{a}_{k}^{\dagger} \ket{g,vac}   \notag \\
    && +\beta(t) \sum_{m=1}^{N} \int^{+\infty}_{-\infty} g_{m} \sqrt{|k|} e^{-ik{x_{m}}} \hat{a}_{k}^{\dagger} \ket{g,vac} \mathrm{d}k + \sum_{m=1}^{N} \int^{+\infty}_{-\infty} g_{m} \sqrt{|k|} e^{ik{x_{m}}} \alpha_{k}(t)  \ket{e,vac} \mathrm{d}k    \notag \\
    &=& i\hbar \dfrac{\mathrm{d}}{\mathrm{d}t} \beta(t) \ket{e,vac} + i\hbar \int \mathrm{d}k \dfrac{\mathrm{d}}{\mathrm{d}t} \alpha_{k}(t) \hat{a}_{k}^{\dagger} \ket{g,vac}.
\eea
Therefore, the dynamics for the giant atom is given by
\bea \label{expression of d_beta(t)}
    \dfrac{\mathrm{d}}{\mathrm{d}t} \beta(t) = -i\Omega \beta(t) - \sum_{m=1}^{N} \int^{+\infty}_{-\infty} \dfrac{i}{\hbar} g_{m} \sqrt{|k|} e^{ik{x_{m}}} \alpha_{k}(t) \mathrm{d}k,
\eea
and the dynamics for the propagating modes in the waveguide is
\bea \label{expression of d_alpha}
    \dfrac{\mathrm{d}}{\mathrm{d}t} \alpha_{k}(t) = -i\omega_{k} \alpha_{k}(t) - \beta(t) \sum_{m=1}^{N} \dfrac{i}{\hbar} g_{m} \sqrt{|k|} e^{-ik{x_{m}}}.
\eea
The formal solution of Eq.~(\ref{expression of d_alpha}) is given by
\bea \label{expression of alpha}
    \alpha_{k}(t) = e^{-i\omega_{k}t} \left\{ \alpha_{k}(0) - \sum_{m=1}^{N} \dfrac{i}{\hbar} g_{m} \sqrt{|k|} e^{-ik{x_{m}}} \int_{0}^{t} \beta(t') e^{i\omega_{k} t'} \mathrm{d}t' \right\}.
\eea
Inserting Eq.~(\ref{expression of alpha}) into Eq.~(\ref{expression of d_beta(t)}), we obtain
\bea \label{expression of d_beta(t)_half}
    \dfrac{\mathrm{d}}{\mathrm{d}t} \beta(t) &=& -i\Omega \beta(t) - \sum_{m=1}^{N} \int^{+\infty}_{-\infty} \dfrac{i}{\hbar} g_{m} \sqrt{|k|} e^{i(k{x_{m}}-\omega_{k} t)} \alpha_{k}(0) \mathrm{d}k \notag \\
    &&- \dfrac{1}{\hbar^{2}} \sum_{m,m'=1}^{N} g_{m} g_{m'} \int_{0}^{t} \beta(t') \mathrm{d}t' \int^{+\infty}_{-\infty} |k| e^{ik(x_{m}-x_{m'})+i\omega_{k}(t'-t)} \mathrm{d}k.
\eea
Using the relation $\omega_{k} = |k|v$, the third term of right-hand side of Eq.~(\ref{expression of d_beta(t)_half}) can be simplified as
\bea \label{third term}
    && \int^{+\infty}_{-\infty} |k| e^{ik(x_{m}-x_{m'})+i\omega_{k}(t'-t)} \mathrm{d}k \notag \\
    && = \int^{0}_{-\infty} \dfrac{\omega_{k}}{v} e^{i(-\omega_{k}/v)(x_{m}-x_{m'})+i\omega_{k}(t'-t)} \mathrm{d} \left(\dfrac{-\omega_{k}}{v} \right) + \int^{+\infty}_{0} \dfrac{\omega_{k}}{v} e^{i(\omega_{k}/v)(x_{m}-x_{m'})+i\omega_{k}(t'-t)} \mathrm{d}\left(\dfrac{\omega_{k}}{v} \right) \notag \\
    && = \dfrac{1}{v^{2}} \int^{+\infty}_{0} \omega_{k} \left[e^{i\omega_{k} (x_{m}-x_{m'}) /v + i\omega_{k} (t'-t)} + e^{i\omega_{k} (x_{m'}-x_{m}) /v + i\omega_{k} (t'-t)} \right] \mathrm{d}\omega_{k}.
\eea
Inserting Eq.~(\ref{third term}) into Eq.~(\ref{expression of d_beta(t)_half}), we can derive,
\bea \label{dbeta(t)-middle}
    \dfrac{\mathrm{d}}{\mathrm{d}t} \beta(t) &= 
    & -i\Omega \beta(t) - \sum_{m=1}^{N} \int^{+\infty}_{-\infty} \dfrac{i}{\hbar} g_{m} \sqrt{|k|} e^{i(k{x_{m}}-\omega_{k} t)} \alpha_{k}(0) \mathrm{d}k - \dfrac{1}{\hbar^{2} v^{2}} \sum_{m,m'=1}^{N} g_{m} g_{m'} \int_{0}^{t} \beta(t') \mathrm{d}t' \notag \\
    && \times \int^{+\infty}_{0} \omega_{k}  \left[e^{i\omega_{k} (x_{m}-x_{m'}) /v + i\omega_{k} (t'-t)} + e^{i\omega_{k} (x_{m'}-x_{m}) /v + i\omega_{k} (t'-t)} \right] \mathrm{d}\omega_{k}.
\eea

To further simplify Eq.~(\ref{dbeta(t)-middle}), we adopt the Weisskopf-Wigner approximation. In the emission spectrum, the intensity of the emitted radiation is concentrated in the range around the atomic transition frequency $\Omega$. By setting $g_{m} = G_{m} g_{0}$, $g_{m'} = G_{m'} g_{0}$ and $\gamma = \dfrac{4\pi g_0^2 \Omega}{\hbar^2 v^2}$, we have
\bea \label{before-laplace}
    \dfrac{\mathrm{d}}{\mathrm{d}t}\beta(t)&\approx 
    & -i\Omega\beta(t)-\dfrac{\gamma}{2}\sum^N_{m,m^{'}=1}G_{m}G_{m^{'}} \int_{0}^{t} \beta(t') \mathrm{d}t' [ \delta(t'-t+\tau_{mm'}) + \delta(t'-t-\tau_{mm'}) ] \notag \\
    && -i\sqrt{\dfrac{\gamma v}{4\pi}}\sum^{N}_{m=1}\int^{+\infty}_{-\infty}G_{m}e^{i(kx_{m}-\omega_{k}t)}\alpha_{k}(0)\mathrm{d}k \notag \\
    &=&-i\Omega\beta(t)-\dfrac{\gamma}{2}\sum^N_{m,m^{'}=1}G_{m}G_{m^{'}}\beta(t-|\tau_{mm'}|)\Theta(t-|\tau_{mm'}|) \notag \\
    &&- i\sqrt{\dfrac{\gamma v}{4\pi}}\sum^{N}_{m=1}\int^{+\infty}_{-\infty}G_{m}e^{i(kx_{m}-\omega_{k}t)}\alpha_{k}(0)\mathrm{d}k,
\eea
which is Eq.~(\ref{formal d_beta(t)}) in the main text. Here, we have used $\int^{+\infty}_{-\infty} \mathrm{d}\omega_{k} e^{i\omega_{k}t} = 2\pi\delta(t)$ and $\int^{+\infty}_{-\infty} f(x) \delta(x-x_{0}) \mathrm{d}x = f(x_{0})$.  We have also introduced the delay time $\tau_{mm'} = (x_{m}-x_{m'})/v$, where $\Theta(x)$ is the Heaviside step function defined via $\Theta(x) = 0$ for $x < 0$ and $\Theta(x) = 1$ for $x > 0$. We have made the Markovian approximation at each single coupling point, but retain the time delay (non-Markovian dynamics) between different coupling points.
In order to solve Eq.~(\ref{before-laplace}), we apply the Laplace transformation $E_{\beta}(S) \equiv \int_0^{\infty} \mathrm{d}t \beta(t) e^{-st}$ and obtain
\begin{equation}
    sE_{\beta}(s)-\beta(0) = -i\Omega E_{\beta}(s)-\dfrac{\gamma}{2}\sum^N_{m,m^{'}=1}G_{m}G_{m^{'}}e^{-|\tau_{mm'}|s} E_{\beta}(s) -i\sqrt{\dfrac{\gamma v}{4\pi}}\sum^{N}_{m=1}\int^{+\infty}_{-\infty}G_{m}\dfrac{e^{ikx_{m}}\alpha_{k}(0)}{s+i\omega_{k}}\mathrm{d}k.
\end{equation}
Thus, $E_{\beta}(s)$ can be obtained from the above equation
\bea
    E_{\beta}(s) &= &\dfrac{\beta(0)}{s+i\Omega +\dfrac{\gamma}{2}\sum^N_{m,m^{'}=1}G_{m}G_{m^{'}}e^{-|\tau_{mm'}|s}} \notag \\
    &&-i\sqrt{\dfrac{\gamma v}{4\pi}}\sum^{N}_{m=1}\int^{+\infty}_{-\infty}G_{m}\dfrac{e^{ikx_{m}}\alpha_{k}(0)}{(s+i\omega_{k}) (s+i\Omega +\dfrac{\gamma}{2}\sum^N_{m,m^{'}=1}G_{m}G_{m^{'}}e^{-|\tau_{mm'}|s})}\mathrm{d}k.
\eea
The poles of $E_{\beta}(s)$ are $s_{\rm{pole}} = -i \omega_k$ and are also given by the roots of the following equation:
\begin{equation} \label{equation of pole}
    s_n+i\Omega +\dfrac{\gamma}{2}\sum^N_{m,m^{'}=1}G_{m}G_{m^{'}}e^{-|\tau_{mm'}|s_n}=0.
\end{equation}
The time evolution of $\beta(t)$ can be obtained by the inverse Laplace transformation,
\begin{equation}
    \beta(t)=\sum e^{s_{pole}t} {\lim_{s\to s_{pole}}}E_{\beta}(s)(s-s_{pole}),
\end{equation}
thus
\bea
    \beta(t)&=&{\lim_{s\to s_n}}\sum_n \dfrac{\beta(0)e^{{s_n}t}(s-s_n)}{s+i\Omega +\dfrac{\gamma}{2}\sum^N_{m,m^{'}=1}G_{m}G_{m^{'}}e^{-|\tau_{mm'}|s}} \notag \\
    &&-i\sqrt{\dfrac{\gamma v}{4\pi}}{\lim_{s\to {-i\omega_{k}} }}\sum^{N}_{m=1}\int^{+\infty}_{-\infty}G_{m}\dfrac{e^{ikx_{m}}\alpha_{k}(0)e^{-i\omega_{k}t}(s+i\omega_{k})}{(s+i\omega_{k}) (s+i\Omega +\dfrac{\gamma}{2}\sum^N_{m,m^{'}=1}G_{m}G_{m^{'}}e^{-|\tau_{mm'}|s})}\mathrm{d}k \notag \\
    &&-i\sqrt{\dfrac{\gamma v}{4\pi}}{\lim_{s\to {s_n} }}\sum^{N}_{m=1}\sum_n\int^{+\infty}_{-\infty}G_{m}\dfrac{e^{ikx_{m}}\alpha_{k}(0)e^{{s_n}t}(s-s_n)}{(s+i\omega_{k}) (s+i\Omega +\dfrac{\gamma}{2}\sum^N_{m,m^{'}=1}G_{m}G_{m^{'}}e^{-|\tau_{mm'}|s})}\mathrm{d}k.\nn
\eea
According to the L'H$\mathrm{\hat{o}}$pital rule and Eq.~(\ref{equation of pole}), we have
\begin{align}
    \beta(t)=&\sum_n \dfrac{\beta(0)e^{{s_n}t}}{1-\dfrac{\gamma}{2}\sum^N_{m,m^{'}=1}G_{m}G_{m^{'}}e^{-|\tau_{mm'}|s_n} |\tau_{mm'}|} \notag \\
    &-i\sqrt{\dfrac{\gamma v}{4\pi}}\sum^{N}_{m=1}\int^{+\infty}_{-\infty}G_{m}\dfrac{e^{ikx_{m}-i\omega_{k}t}\alpha_{k}(0)}{i(\Omega-\omega_{k})+\dfrac{\gamma}{2}\sum^N_{m,m^{'}=1}G_{m}G_{m^{'}}e^{i\omega_{k}|\tau_{mm'}|}}\mathrm{d}k \notag \\
    &-i\sqrt{\dfrac{\gamma v}{4\pi}}\sum^{N}_{m=1}\sum_n\int^{+\infty}_{-\infty}G_{m}\dfrac{e^{ikx_{m}+{s_n}t}\alpha_{k}(0)}{(s_n+i\omega_{k}) (1-\dfrac{\gamma}{2}\sum^N_{m,m^{'}=1}G_{m}G_{m^{'}}e^{-|\tau_{mm'}|s_n} |\tau_{mm'}|)}\mathrm{d}k.
\end{align}
For the spontaneous emission ($\beta(0) = 1$, $\alpha_k(0) = 0$), the above equation becomes Eq.~(\ref{eq-laplace}) in the main text.

Since each $\alpha_{k}(t)$ represents the time-dependent probability amplitude of a plane wave $e^{ikx}$, the total time-dependent field function in the waveguide is given by
\bea
    \Phi(x,t) \equiv \dfrac{1}{\sqrt{2 \pi}} \int^{+\infty}_{-\infty} e^{ikx} \alpha_{k}(t) \mathrm{d}k.
\eea
From Eq.~(\ref{expression of alpha}), we have
\bea
\Phi(x,t) &=& \dfrac{1}{\sqrt{2 \pi}} \int^{+\infty}_{-\infty} e^{i(kx-\omega_{k}t)} \alpha_{k}(0) \mathrm{d}k - i \dfrac{1}{\sqrt{2 \pi}} \sqrt{\dfrac{\gamma v}{4\pi}}\sum^{N}_{m=1} \int_{0}^{t} \beta(t') \mathrm{d}t' \int^{+\infty}_{-\infty} e^{ik(x - x_{m})+i\omega_{k}(t'-t)} \mathrm{d}k \notag \\
&\approx& \dfrac{1}{\sqrt{2 \pi}} \int^{+\infty}_{-\infty} e^{i(kx-\omega_{k}t)} \alpha_{k}(0) \mathrm{d}k 
- i \sqrt{\dfrac{\gamma}{2v}} \sum^{N}_{m=1} G_{m} \beta \left( t - \frac{|x - x_m|}{v} \right) \Theta \left( t - \frac{|x - x_m|}{v} \right),\ \ \ \  \ \ \ \ \ 
\eea
which is Eq.~(\ref{bosonic field}) in the main text. In the last step, we have used the Weisskopf-Wigner approximation again as we did in the derivation of Eq.~(\ref{before-laplace}).

\section{Derivation for the vectorized density matrix} \label{Derive the vectorized density matrix}

We rewrite Eq.~(\ref{master-equation}) in the main text,
\bea \label{master-equation-appendix}
    \dot{\rho} = &-&i \left[ \omega_a^{'} \dfrac{\sigma_{z}^{a}}{2} + \omega_b^{'} \dfrac{\sigma_{z}^{b}}{2} + f ( \sigma_{-}^{a} \sigma_{+}^{b} + \sigma_{+}^{a} \sigma_{-}^{b}) , \rho \right] + \Gamma_a \mathcal{D} [\sigma_{-}^{a}] \rho + \Gamma_b \mathcal{D} [\sigma_{-}^{b}] \rho \notag \\
    &+& \Gamma_{\rm{coll}} \left\{ \sigma_{-}^{a} \rho \sigma_{+}^{b} + \sigma_{-}^{b} \rho \sigma_{+}^{a} - \dfrac{1}{2} [ ( \sigma_{+}^{a} \sigma_{-}^{b} + \sigma_{+}^{b} \sigma_{-}^{a} ) \rho + \rho ( \sigma_{+}^{a} \sigma_{-}^{b} + \sigma_{+}^{b} \sigma_{-}^{a} )] \right\}.
\eea
We now choose a set of basis vectors by $\ket{gg} = \ket{1},\ket{ge} = \ket{2},\ket{eg} = \ket{3},\ket{ee} = \ket{4}$ and vectorize the density matrix via $\vec{\rho}\equiv (\bra{1} \rho \ket{1}, \cdots, \bra{q} \rho \ket{j},\cdots,\bra{4} \rho \ket{4})^T$ with $q,j = 1,2,3,4$.
Using these four basis on both sides of Eq.~(\ref{master-equation-appendix}), we derive the dynamics for the elements of the vectorized density matrix $\bra{q} \dot{\rho} \ket{j}(q,j = 1,2,3,4)$ on the left-hand side of Eq.~(\ref{master-equation-appendix}). The right-hand side of Eq.~(\ref{master-equation-appendix}) can be split into four terms as follows.
The first term is given by
\bea
    -&& \bra{q} \left[ i \dfrac{\omega_a^{'}}{2} \sigma_{z}^{a} + i \dfrac{\omega_b^{'}}{2} \sigma_{z}^{b} + i f ( \sigma_{-}^{a} \sigma_{+}^{b} + \sigma_{+}^{a} \sigma_{-}^{b}) \right] \rho - \rho \left [ i \dfrac{\omega_a^{'}}{2} \sigma_{z}^{a} + i \dfrac{\omega_b^{'}}{2} \sigma_{z}^{b} + i f ( \sigma_{-}^{a} \sigma_{+}^{b} + \sigma_{+}^{a} \sigma_{-}^{b}) \right] \ket{j} \notag \\
    = &&- i \dfrac{\omega_a^{'}}{2} \sum_{n} \bra{q} \sigma_{z}^{a} \ket{n} \bra{n} \rho \ket{j} - i \dfrac{\omega_b^{'}}{2} \sum_{n} \bra{q} \sigma_{z}^{b} \ket{n} \bra{n} \rho \ket{j} - i f \sum_{n} \bra{q} \sigma_{-}^{a} \sigma_{+}^{b} {\ket{n}} \bra{n} \rho \ket{j} \notag \\
    && - i f \sum_{n} \bra{q} \sigma_{+}^{a} \sigma_{-}^{b} {\ket{n}} \bra{n} \rho \ket{j} + i \dfrac{\omega_a^{'}}{2} \sum_{n} \bra{q} \rho \ket{n} \bra{n} \sigma_{z}^{a} \ket{j} + i \dfrac{\omega_b^{'}}{2} \sum_{n} \bra{q} \rho \ket{n} \bra{n} \sigma_{z}^{b} \ket{j} \notag \\
    && + i f \sum_{n} \bra{q} \rho \ket{n} \bra{n} \sigma_{-}^{a} \sigma_{+}^{b} \ket{j} + i f \sum_{n} \bra{q} \rho \ket{n} \bra{n} \sigma_{+}^{a} \sigma_{-}^{b} \ket{j} \notag \\
    = && - i \dfrac{\omega_a^{'}}{2} \sum_{n} \bra{q} \sigma_{z}^{a} \ket{n} \rho_{nj} - i \dfrac{\omega_b^{'}}{2} \sum_{n} \bra{q} \sigma_{z}^{b} \ket{n} \rho_{nj} - i f \sum_{n} \bra{q} \sigma_{-}^{a} \sigma_{+}^{b} {\ket{n}} \rho_{nj} - i f \sum_{n} \bra{q} \sigma_{+}^{a} \sigma_{-}^{b} {\ket{n}} \rho_{nj} \notag \\
    && + i \dfrac{\omega_a^{'}}{2} \sum_{n} \bra{n} \sigma_{z}^{a} \ket{j} \rho_{qn} + i \dfrac{\omega_b^{'}}{2} \sum_{n} \bra{n} \sigma_{z}^{b} \ket{j} \rho_{qn} + i f \sum_{n} \bra{n} \sigma_{-}^{a} \sigma_{+}^{b} \ket{j} \rho_{qn} + i f \sum_{n} \bra{n} \sigma_{+}^{a} \sigma_{-}^{b} \ket{j} \rho_{qn} \notag \\
    = && - i \dfrac{\omega_a^{'}}{2} \sum_{k,m,n} \bra{q} \sigma_{z}^{a} \ket{n} \delta_{nk} \delta_{jm} \rho_{km} - i \dfrac{\omega_b^{'}}{2} \sum_{k,m,n} \bra{q} \sigma_{z}^{b} \ket{n} \delta_{nk} \delta_{jm} \rho_{km} - i f \sum_{k,m,n} \bra{q} \sigma_{-}^{a} \sigma_{+}^{b} {\ket{n}} \delta_{nk} \delta_{jm} \rho_{km} \notag \\
    &&- i f \sum_{k,m,n} \bra{q} \sigma_{+}^{a} \sigma_{-}^{b} {\ket{n}} \delta_{nk} \delta_{jm} \rho_{km} + i \dfrac{\omega_a^{'}}{2} \sum_{k,m,n} \bra{n} \sigma_{z}^{a} \ket{j} \delta_{qk} \delta_{nm} \rho_{km} + i \dfrac{\omega_b^{'}}{2} \sum_{k,m,n} \bra{n} \sigma_{z}^{b} \ket{j} \delta_{qk} \delta_{nm} \rho_{km} \notag \\
    && + i f \sum_{k,m,n} \bra{n} \sigma_{-}^{a} \sigma_{+}^{b} \ket{j} \delta_{qk} \delta_{nm} \rho_{km} + i f \sum_{k,m,n} \bra{n} \sigma_{+}^{a} \sigma_{-}^{b} \ket{j} \delta_{qk} \delta_{nm} \rho_{km},
\eea
where $k,m,n = 1,2,3,4$. 
The second term is given by
\bea
    && \bra{q} \Gamma_a \mathcal{D} [\sigma_{-}^{a}] \rho \ket{j} \notag \\
    &&= \bra{q} \Gamma_a \left( \sigma_{-}^{a} \rho \sigma_{+}^{a} - \dfrac{1}{2} \sigma_{+}^{a}\sigma_{-}^{a} \rho - \dfrac{1}{2} \rho \sigma_{+}^{a} \sigma_{-}^{a} \right) \ket{j} = \Gamma_a \bra{q} \sigma_{-}^{a} \rho \sigma_{+}^{a} \ket{j} - \dfrac{1}{2} \Gamma_a \bra{q} \sigma_{+}^{a} \sigma_{-}^{a} \rho \ket{j} - \dfrac{1}{2} \Gamma_a \bra{q} \rho \sigma_{+}^{a} \sigma_{-}^{a} \ket{j} \notag \\
    &&= \Gamma_a \sum_{n,p} \bra{q} \sigma_{-}^{a} \ket{n} \bra{n} \rho \ket{p} \bra{p} \sigma_{+}^{a} \ket{j} - \dfrac{1}{2} \Gamma_a \sum_{n} \bra{q} \sigma_{+}^{a} \sigma_{-}^{a} \ket{n} \bra{n} \rho \ket{j} - \dfrac{1}{2} \Gamma_a \sum_{n} \bra{q} \rho \ket{n} \bra{n} \sigma_{+}^{a} \sigma_{-}^{a} \ket{j} \notag \\
    &&= \Gamma_a \sum_{k,m,n,p} \bra{q} \sigma_{-}^{a} \ket{n} \bra{p} \sigma_{+}^{a} \ket{j} \delta_{nk} \delta_{pm} \rho_{km} - \dfrac{1}{2} \Gamma_a \sum_{k,m,n} \bra{q} \sigma_{+}^{a} \sigma_{-}^{a} \ket{n} \delta_{nk} \delta_{jm} \rho_{km} \notag \\
    &&\quad - \dfrac{1}{2} \Gamma_a \sum_{k,m,n} \bra{n} \sigma_{+}^{a} \sigma_{-}^{a} \ket{j} \delta_{qk} \delta_{nm} \rho_{km},
\eea
where $p = 1,2,3,4$. The third term is given by
\bea
    \bra{q} \Gamma_b \mathcal{D} [\sigma_{-}^{b}] \rho \ket{j} &=& \Gamma_b \sum_{k,m,n,p} \bra{q} \sigma_{-}^{b} \ket{n} \bra{p} \sigma_{+}^{b} \ket{j} \delta_{nk} \delta_{pm} \rho_{km} - \dfrac{1}{2} \Gamma_b \sum_{k,m,n} \bra{q} \sigma_{+}^{b} \sigma_{-}^{b} \ket{n} \delta_{nk} \delta_{jm} \rho_{km} \notag \\
    && - \dfrac{1}{2} \Gamma_b \sum_{k,m,n} \bra{n} \sigma_{+}^{b} \sigma_{-}^{b} \ket{j} \delta_{qk} \delta_{nm} \rho_{km}.
\eea
The last term is given by
\bea
    && \bra{q} \Gamma_{\rm{coll}} \left\{ \sigma_{-}^{a} \rho \sigma_{+}^{b} + \sigma_{-}^{b} \rho \sigma_{+}^{a} - \dfrac{1}{2} [ ( \sigma_{+}^{a} \sigma_{-}^{b} + \sigma_{+}^{b} \sigma_{-}^{a} ) \rho + \rho ( \sigma_{+}^{a} \sigma_{-}^{b} + \sigma_{+}^{b} \sigma_{-}^{a} )] \right\} \ket{j} \notag \\
    &&= \Gamma_{\rm{coll}} \sum_{k,m,n,p} \bra{q} \sigma_{-}^{a} \ket{n} \bra{p} \sigma_{+}^{b} \ket{j} \delta_{nk} \delta_{pm} \rho_{km} + \Gamma_{\rm{coll}} \sum_{k,m,n,p} \bra{q} \sigma_{-}^{b} \ket{n} \bra{p} \sigma_{+}^{a} \ket{j} \delta_{nk} \delta_{pm} \rho_{km} \notag \\
    &&\quad - \dfrac{1}{2} \Gamma_{\rm{coll}} \sum_{k,m,n} \bra{q} \sigma_{+}^{a} \sigma_{-}^{b} \ket{n} \delta_{nk} \delta_{jm} \rho_{km} - \dfrac{1}{2} \Gamma_{\rm{coll}} \sum_{k,m,n} \bra{q} \sigma_{+}^{b} \sigma_{-}^{a} \ket{n} \delta_{nk} \delta_{jm} \rho_{km} \notag \\
    &&\quad - \dfrac{1}{2} \Gamma_{\rm{coll}} \sum_{k,m,n} \bra{n} \sigma_{+}^{a} \sigma_{-}^{b} \ket{j} \delta_{qk} \delta_{nm} \rho_{km} - \dfrac{1}{2} \Gamma_{\rm{coll}} \sum_{k,m,n} \bra{n} \sigma_{+}^{b} \sigma_{-}^{a} \ket{j} \delta_{qk} \delta_{nm} \rho_{km}.
\eea
Combining four terms, we have the final expression for the dynamics of the vectorized density matrix
\bea \label{density-matrix-final}
\dot{\rho}_{qj} &= \sum_{k,m}\mathcal{L}_{qj,km} \rho_{km},
\eea
where
\bea
    \mathcal{L}_{qj,km} && =  \sum_{n,p} ( - i \dfrac{\omega_a^{'}}{2} \bra{q} \sigma_{z}^{a} \ket{n} \delta_{nk} \delta_{jm} - i \dfrac{\omega_b^{'}}{2} \bra{q} \sigma_{z}^{b} \ket{n} \delta_{nk} \delta_{jm} - i f \bra{q} \sigma_{-}^{a} \sigma_{+}^{b} {\ket{n}} \delta_{nk} \delta_{jm} - i f \bra{q} \sigma_{+}^{a} \sigma_{-}^{b} {\ket{n}} \delta_{nk} \delta_{jm} \notag \\
    &&\quad + i \dfrac{\omega_a^{'}}{2} \bra{n} \sigma_{z}^{a} \ket{j} \delta_{qk} \delta_{nm} + i \dfrac{\omega_b^{'}}{2} \bra{n} \sigma_{z}^{b} \ket{j} \delta_{qk} \delta_{nm} + i f \bra{n} \sigma_{-}^{a} \sigma_{+}^{b} \ket{j} \delta_{qk} \delta_{nm} + i f \bra{n} \sigma_{+}^{a} \sigma_{-}^{b} \ket{j} \delta_{qk} \delta_{nm} \notag \\
    &&\quad + \Gamma_a \bra{q} \sigma_{-}^{a} \ket{n} \bra{p} \sigma_{+}^{a} \ket{j} \delta_{nk} \delta_{pm} - \dfrac{1}{2} \Gamma_a \bra{q} \sigma_{+}^{a} \sigma_{-}^{a} \ket{n} \delta_{nk} \delta_{jm} - \dfrac{1}{2} \Gamma_a \bra{n} \sigma_{+}^{a} \sigma_{-}^{a} \ket{j} \delta_{qk} \delta_{nm} \notag \\
    &&\quad + \Gamma_b \bra{q} \sigma_{-}^{b} \ket{n} \bra{p} \sigma_{+}^{b} \ket{j} \delta_{nk} \delta_{pm} - \dfrac{1}{2} \Gamma_b \bra{q} \sigma_{+}^{b} \sigma_{-}^{b} \ket{n} \delta_{nk} \delta_{jm} - \dfrac{1}{2} \Gamma_b \bra{n} \sigma_{+}^{b} \sigma_{-}^{b} \ket{j} \delta_{qk} \delta_{nm} \notag \\
    &&\quad + \Gamma_{\rm{coll}} \bra{q} \sigma_{-}^{a} \ket{n} \bra{p} \sigma_{+}^{b} \ket{j} \delta_{nk} \delta_{pm} + \Gamma_{\rm{coll}} \bra{q} \sigma_{-}^{b} \ket{n} \bra{p} \sigma_{+}^{a} \ket{j} \delta_{nk} \delta_{pm} - \dfrac{1}{2} \Gamma_{\rm{coll}} \bra{q} \sigma_{+}^{a} \sigma_{-}^{b} \ket{n} \delta_{nk} \delta_{jm} \notag \\
    &&\quad - \dfrac{1}{2} \Gamma_{\rm{coll}} \bra{q} \sigma_{+}^{b} \sigma_{-}^{a} \ket{n} \delta_{nk} \delta_{jm} - \dfrac{1}{2} \Gamma_{\rm{coll}} \bra{n} \sigma_{+}^{a} \sigma_{-}^{b} \ket{j} \delta_{qk} \delta_{nm} - \dfrac{1}{2} \Gamma_{\rm{coll}} \bra{n} \sigma_{+}^{b} \sigma_{-}^{a} \ket{j} \delta_{qk} \delta_{nm} ).
\eea
The left-hand side of Eq.~(\ref{density-matrix-final}) can be expressed as $\dfrac{\mathrm{d}\vec{\rho}}{\mathrm{d}t}$. As each $\dot{\rho}_{qj}$ can be expressed as the linear combination of $\rho_{km}$, the right-hand side of Eq.~(\ref{density-matrix-final}) can be expressed as $\mathcal{L} \vec{\rho}$, where superoperator $\mathcal{L}$ is a $16\times 16$ coefficient matrix given by Eq.~(\ref{eq-L}) in the main text.

\twocolumngrid

%

\end{document}